\documentclass[widetext,preprint,superscriptaddress,reprint,floatfix,nofootinbib]{revtex4-1}

\expandafter\let\csname equation*\endcsname\relax
\expandafter\let\csname endequation*\endcsname\relax

\usepackage{bm}				
\usepackage{amsmath}
\usepackage{graphicx}
\usepackage{caption}
\usepackage{bbold}

\newcommand{\ket}[1]{\vert{#1}\rangle} 
\newcommand{\bra}[1]{\langle{#1}\vert} 
\newcommand{\bracket}[2]{\langle{#1}\vert{#2}\rangle}

\newcommand{\kket}[1]{\vert{#1}\rangle\rangle} 
\newcommand{\bbra}[1]{\langle\langle{#1}\vert} 
\newcommand{\bbracket}[2]{\langle\langle{#1}\vert{#2}\rangle\rangle} 
\newcommand{\bproj}[2]{\kket{#1}\!\bbra{#1}} 

\newcommand{\proj}[1]{\ket{#1}\!\bra{#1}}
\newcommand{\op}[2]{\ket{#1}\!\bra{#2}}
\newcommand{\mean}[1]{\langle #1 \rangle}

\newcommand{\abs}[1]{\left|#1\right|} 
\newcommand{\norm}[1] {\left| \left| #1\right|\right|} 
\newcommand{\pare}[1]{\left( #1 \right)}
\newcommand{\be}{\begin{equation}}
\newcommand{\ee}{\end{equation}}
\newcommand{\key}[1]{\left\{ #1 \right\}}
\newcommand{\cor}[1]{\left[ #1 \right]}

\newcommand{\bc}{\begin{center}}
\newcommand{\ec}{\end{center}}
\newcommand{\ben}{\begin{eqnarray}}
\newcommand{\een}{\end{eqnarray}}

\newcommand{\cH}{{\cal H}}

\newcommand{\cL}{{\cal L}}

\newcommand{\cV}{{\cal V}}

\newcommand{\id}{\mathbb 1}

\newcommand{\cB}{{\cal B}}

\newcommand{\Tr}{\textrm{Tr}}

\newtheorem{postulate}{Postulate}
\newtheorem{corollary}{Corollary}
\newtheorem{definition}{Definition}
\newtheorem{lemma}{Lemma}
\newtheorem{theorem}{Theorem}

\begin{document}

\title[]{A short introduction to the Lindblad Master Equation}

\author{Daniel Manzano}
\address{Electromagnetism and Condensed Matter Department and Carlos I Institute for Theoretical and Computational Physics. University of Granada. E-18071 Granada. Spain }
\email{manzano@onsager.ugr.es}

\begin{abstract}
The theory of open quantum system is one of the most essential tools for the development of quantum technologies. Furthermore, the Lindblad (or Gorini-Kossakowski-Sudarshan-Lindblad) Master Equation plays a key role as it is the most general generator of Markovian dynamics in quantum systems. In this paper, we present this equation together with its derivation and methods of resolution.  The presentation tries to be as self-contained and straightforward as possible to be useful to readers with no previous knowledge of this field.
\end{abstract}

\onecolumngrid

\maketitle

\onecolumngrid
\section{Introduction}

Open quantum system techniques are vital for many studies in quantum mechanics \cite{gardiner_00,breuer_02,rivas_12}. This happens because closed  quantum systems are just an idealisation of real systems\footnote{The same happens with closed classical systems.}, as in Nature nothing can be isolated. In  practical problems, the interaction of the system of interest with the environment cannot be avoided, and we require an approach in which the environment can be effectively removed from the equations of motion. 

\begin{figure}[h!]
\begin{center}
\includegraphics[scale=0.2]{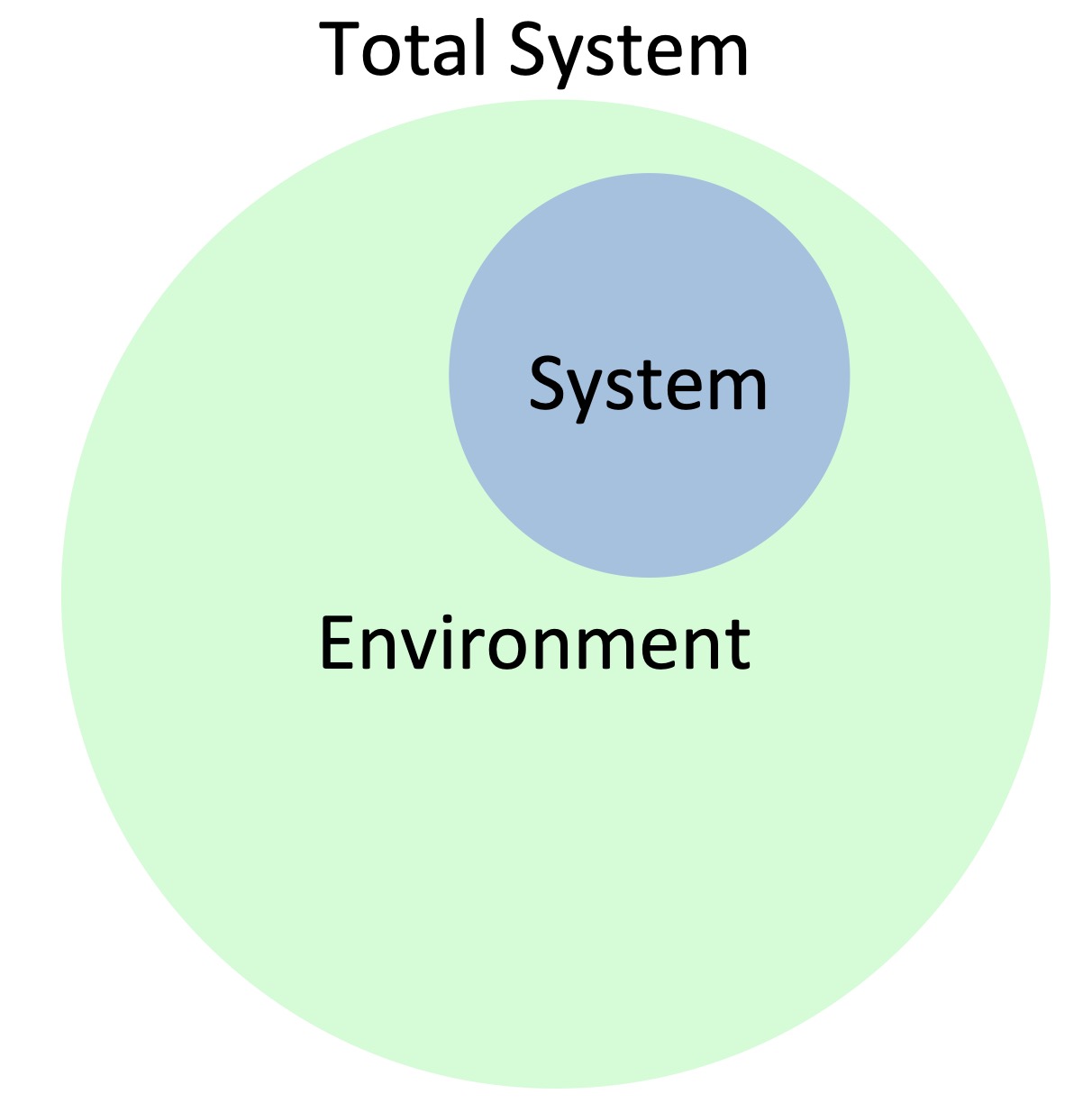}
\end{center}
\caption{A total system divided into the system of interest, ``System'', and the environment.   }
\label{fig:fig0}
\end{figure}

The general problem addressed by Open Quantum Theory is sketched in Figure \ref{fig:fig0}. In the most general picture, we have a total system that conforms a closed quantum system by itself. We are mostly interested in a subsystem of the total one (we call it just ``system'' instead ``total system''). Therefore, the whole system is divided into our system of interest and an environment. The goal of Open Quantum Theory is to infer the equations of motions of the reduced systems from the equation of motion of the total system. For practical purposes, the reduced equations of motion should be easier to solve than the full dynamics of the system. Because of his requirement, several approximations are usually made in the derivation of the reduced dynamics.

One particular, and interesting, case of study is the dynamics of a system connected to several baths modelled by a Markovian interaction. In this case the most general quantum dynamics is generated by the Lindblad equation (also called   Gorini-Kossakowski-Sudarshan-Lindblad equation) \cite{lindblad:cmp76,gorini:jmp76}. It is difficult to overemphasize the importance of this Master Equation. It plays an important role in fields as quantum optics \cite{gardiner_00,manzano:sr16}, condensed matter \cite{prosen:prl11,manzano:pre12,manzano:njp16,olmos:prl12}, atomic physics \cite{metz:prl06,jones:pra18}, quantum information \cite{lidar:prl98,kraus:08}, decoherence \cite{brun:pra00,schlosshauer_07}, and quantum biology \cite{plenio:njp08,mohseni:jcp08, manzano:po13}.

The purpose of this paper is to provide basic knowledge about the Lindblad Master Equation. In Section \ref{sec:math}, the mathematical requirements are introduced while in Section \ref{sec:qm} there is a brief review of quantum mechanical concepts that are required to understand the paper. Section \ref{sec:fl}, includes a description of a mathematical framework, the Fock-Liouville space, that is especially useful to work in this problem. In Section \ref{cpt}, we define the concept of CPT-Maps, derive the Lindblad Master Equation from two different approaches, and we discus several properties of the equation. Finally, Section \ref{sec:resolution} is devoted to the resolution of the master equation using different methods. To deepen in the techniques of solving the Lindblad equation, an example consisting of a two-level system with decay is analysed, illustrating the content of every section. The problems proposed are solved by the use of Mathematica notebooks that can be found at \cite{notebook}.

\section{Mathematical basis}
\label{sec:math}

The primary mathematical tool in quantum mechanics is the theory of Hilbert spaces. This mathematical framework allows extending many results from finite linear vector spaces to infinite ones. In any case, this tutorial deals only with finite systems and, therefore, the expressions `Hilbert space' and `linear space' are equivalent. We assume that the reader is skilled in operating in Hilbert spaces. To deepen in the field of Hilbert spaces we recommend the book by Debnath and Mikusi\'nki \cite{debnath_05}. If the reader  needs a brief review of the main concepts required for understanding this paper, we may recommend Nielsen and Chuang's Quantum Computing book \cite{nielsen_00}. It is also required some basic knowledge about infinitesimal calculus, like integration, derivation, and the resolution of simple differential equations,  To help the readers, we have made a glossary of the most used mathematical terms. It can be used also as a checklist of terms the reader should be familiar with.

\vspace{0.25cm}
\noindent
{\bf Glossary:}

\begin{itemize}
\item $\cH$ represents a Hilbert space, usually the space of pure states of a system.
\item $\ket{\psi}\in \cH$ represents a vector of the Hilbert space $\cH$ (a column vector).
\item $\bra{\psi}\in \cH$ represents a vector of the dual Hilbert space of $\cH$ (a row vector).
\item $\bracket{\psi}{\phi}\in \mathbb{C}$ is the scalar product of vectors $\ket{\psi}$ and $\ket{\phi}$.
\item $\norm{\ket{\psi}}$ is the norm of vector $\ket{\psi}$.  $\norm{\ket{\psi}}\equiv\sqrt{\bracket{\psi}{\psi}}$.
\item $B(\cH)$ represents the space of bounded operators acting on the Hilbert space $B:\cH \to \cH$.
\item $\id_{\cH}\in B(\cH)$ is the Identity Operator of the Hilbert space $\cH$ s.t. $\id_{\cH}\ket{\psi}=\ket{\psi},\; \; \forall \ket{\psi}\in \cH $.
\item $\op{\psi}{\phi}\in B(\cH)$ is the operator such that $\pare{\op{\psi}{\phi}} \ket{\varphi}=\bracket{\phi}{\varphi} \ket{\psi},\; \;  \forall \ket{\varphi} \in \cH$.
\item  $O^\dagger\in B(\cH)$ is the Hermitian conjugate of the operator $O\in B(\cH)$.
\item $U\in B(\cH)$ is a unitary operator iff $U U^{\dagger}=U^{\dagger}U=\id$.
\item $H\in B(\cH)$ is a Hermitian operator iff $H=H^{\dagger}$.
\item $A\in B(\cH)$ is a positive operator $\pare{A> 0}$  iff $\bra{\phi} A \ket{\phi}\ge0,\;\;  \forall \ket{\phi}\in \cH$ 
\item $P\in B(\cH)$ is a proyector iff $P P=P$.
\item $\Tr\cor{B}$ represents the trace of  operator $B$.
\item $\rho\pare{\cL}$ represents the space of density matrices, meaning the space of bounded operators acting on $\cH$ with trace $1$ and positive.
\item $\kket{\rho}$ is a vector in the Fock-Liouville space.
\item $\bbracket{A}{B}=\Tr\cor{A^\dagger B}$ is the scalar product of operators $A,B\in B(\cH)$ in the Fock-Liouville space.
\item $\tilde{\cL}$ is the matrix representation of a superoperator in the Fock-Liouville space.
\end{itemize}

\section{(Very short) Introduction to quantum mechanics}
\label{sec:qm}

The purpose of this chapter is to refresh the main concepts of quantum mechanics necessary to understand the Lindblad Master Equation. Of course, this is NOT a full quantum mechanics course. If a reader has no background in this field, just reading this chapter would be insufficient to understand the remaining of this tutorial. Therefore, if the reader is unsure of his/her capacities, we recommend to go first through a quantum mechanics course or to read an introductory book carefully. There are many great quantum mechanics books in the market. For beginners, we recommend Sakurai's book \cite{sakurai_94} or Nielsen and Chuang's Quantum Computing book \cite{nielsen_00}. For more advanced students, looking for a solid mathematical description of quantum mechanics methods, we recommend Galindo and Pascual \cite{galindo_pascual_90}. Finally, for a more philosophical discussion, you should go to Peres' book \cite{peres_95}.

We start stating the quantum mechanics postulates that we need to understand the derivation and application of the Lindblad Master Equation. The first postulate is related to the concept of a quantum state.

\vspace{0.5cm}
\begin{postulate} 
Associated to any isolated physical system, there is a complex Hilbert space $\cH$, known as the {\bf state space} of the system. The state of the system is entirely described by a {\it state vector}, which is a unit vector of the Hilbert space $(\ket{\psi}\in \cH)$.
\end{postulate}
\vspace{0.5cm}

\noindent
As quantum mechanics is a general theory (or a set of theories), it does not tell us which is the proper Hilbert space for each system. This is usually done system by system.  A natural question to ask is if there is a one-to-one correspondence between unit vectors and physical states, meaning that if every unit vector corresponds to a physical system. This is resolved by the following corollary that is a primary ingredient for quantum computation theory (see Ref. \cite{nielsen_00} Chapter 7). 

\vspace{0.5cm}
\begin{corollary} 
All unit vectors of a finite Hilbert space correspond to possible physical states of a system.
\end{corollary}
\vspace{0.5cm}

\noindent
Unit vectors are also called {\it pure states}. If we know the pure state of a system, we have all physical information about it, and we can calculate the probabilistic outcomes of any potential measurement (see the next postulate). This is a very improbable situation as experimental settings are not perfect, and in most cases, we have only imperfect information about the state. Most generally, we may know that a quantum system can be in one state of a set $\key{\ket{\psi_i}}$ with probabilities $p_i$. Therefore, our knowledge of the system is given by an {\it ensemble of pure states} described by the set $\key{\ket{\psi_i}, \; p_i}$. If more than one $p_i$ is different from zero the state is not pure anymore, and it is called a {\it mixed state}. The mathematical tool that describes our knowledge of the system, in this case, is the {\it density operator} (or {\it density matrix}).

\be
\rho \equiv \sum_i p_i \op{\psi_i}{\psi_i}.
\label{eq:dm}
\ee
Density matrices are bounded operators that fulfil two mathematical conditions 

\begin{enumerate}
\item A density matrix $\rho$ has unit trace $\pare{\Tr[\rho]=1 }$.
\item A density matrix is a positive matrix $\rho>0$.
\end{enumerate}
Any operator fulfilling these two properties is considered a density operator. It can be proved trivially that density matrices are also Hermitian. 

If we are given a density matrix, it is easy to verify if it belongs to a pure or a mixed state. For pure states, and only for them, $\Tr[\rho^2]=\Tr[\rho]=1$. Therefore, if  $\Tr[\rho^2]<1$ the system is mixed. The quantity $\Tr[\rho^2]$ is called the purity of the states, and it fulfils the bounds $\frac{1}{d} \le \Tr[\rho^2] \le 1$, being $d$ the dimension of the Hilbert space.

If we fix an arbitrary basis  $\key{\ket{i}}_{i=1}^N$ of the Hilbert space the density matrix in this basis is written as $\rho=\sum_{i,j=1}^N \rho_{i,j} \op{i}{j}$, or 

\be
\rho=
\begin{pmatrix}
\rho_{00} & \rho_{01} & \cdots & \rho_{0N} \\
\rho_{10} & \rho_{11} & \cdots & \rho_{1N} \\
\vdots & \vdots & \ddots & \vdots \\
\rho_{N0} & \rho_{N1} & \cdots & \rho_{NN}
\end{pmatrix},
\ee
where the diagonal elements are called {\it populations} $\pare{\rho_{ii}\in\mathbb{R}_0^+\text{ and } \sum_{i} \rho_{i,i}=1}$, while the off-diagonal elements are called {\it coherences}  $\pare{ \rho^{\phantom{*}}_{i,j} \in \mathbb{C} \text{ and }  \rho^{\phantom{*}}_{i,j}=\rho_{j,i}^*}$. Note that this notation is base-dependent.

\bc
\vspace{0.5cm}
\framebox[15.5cm][l]{
\begin{minipage}[l]{15cm}
\vspace{0.25cm} 

{\bf Box 1. State of a two-level system (qubit)}
\vspace{0.25cm}

The Hilbert space of a two-level system is just the two-dimension lineal space $\cH_2$. Examples of this kind of system are  $\frac{1}{2}$-spins and two-level atoms. We can define a basis of it by the orthonormal vectors: $\key{ \ket{0},\;\ket{1}}$. A pure state of the system would be any unit vector of $\cH_2$. It can always be expressed as a $\ket{\psi}=a\ket{0} + b\ket {1}$ with $a,b \in \mathbb{C}$ s. t. $\abs{a}^2 + \abs{b}^2=1$. 
\vspace{0.25cm}

A mixed state is therefore represented by a positive unit trace operator  $ \rho\in O(\cH_2)$. 

\be
\rho =
\begin{pmatrix}
\rho_{00} & \rho_{01} \\
\rho_{10} & \rho_{11} 
\end{pmatrix}
= \rho_{00} \proj{0} + \rho_{01} \op{0}{1} + \rho_{10} \op{1}{0} + \rho_{11} \proj{1},
\label{eq:denmat}
\ee
ant it should fulfil $\rho_{00}+\rho_{11}=1$ and $\rho_{01}^{\phantom{*}}=\rho_{10}^*$.
\vspace{0.25cm}
\end{minipage}
\label{minipage1}
}
\ec
\vspace{0.5cm}

\noindent
Once we know the state of a system, it is natural to ask about the possible outcomes of experiments (see Ref. \cite{sakurai_94}, Section 1.4). 

\vspace{0.5cm}
\begin{postulate} 

All possible measurements in a quantum system are described by a Hermitian operator or {\bf observable}.  Due to the Spectral Theorem we know that any observable $O$ has a spectral decomposition in the form\footnote{For simplicity, we assume a non-degenerated spectrum.} 

\be
O=\sum_i a_i \op{a_i}{a_i}, 
\ee
being $a_i\in\mathbb{R}$ the eigenvalues of the observable and $\ket{a_i}$ their corresponding eigenvectors. The probability of obtaining the result $a_i$ when measuring the property described by observable $O$ in a state $\ket{\psi}$ is given by 

\be
P(a_i)= \left|  \bracket{\psi}{a_i}	  \right|^2.
\ee
After the measurement we obtain the state $\ket{a_i}$ if the outcome $a_i$ was measured. This is called the {\it post-measurement state}.
\label{post4}
\end{postulate}

\vspace{0.5cm}

This postulate allow us to calculate the possible outputs of a system, the probability of these outcomes, as well as the after-measurement state. A measurement usually changes the state, as it can only remain unchanged if it was already in an eigenstate of the observable. 

It is possible to calculate the expectation value of the outcome of a measurement defined by operator $O$ in a state $\ket{\psi}$ by just applying the simple formula 

\be
\mean{O}= \bra{\psi} O \ket{\psi}.
\ee
With a little algebra we can translate this postulate to mixed states. In this case, the probability of obtaining an output $a_i$ that corresponds to an eigenvector $\ket{a_i}$ is 

\be
P(a_i)=\Tr\cor{\proj{a_i}\rho},
\ee
and the expectation value of operator $O$ is 

\be
\mean{O}=\Tr \cor{O\rho}.
\ee

\bc
\vspace{0.5cm}
\framebox[15.5cm][l]{
\begin{minipage}[l]{15cm}
\vspace{0.25cm} 

{\bf Box 2. Measurement in a two-level system.}
\vspace{0.25cm}

A possible test to perform in our minimal model is to measure the energetic state of a system, assuming that both states have a different energy. The observable corresponding to this measurement would be 

\be
H=E_0 \proj{0} + E_1 \proj{1}.
\ee

This operator has two eigenvalues $\key{E_0,\;E_1}$ with two corresponding eigenvectors $\key{\ket{0},\; \ket{1}}$. 

\vspace{0.25cm}

If we have a pure state $\psi=a\ket{0} + b \ket{1}$ the probability of measuring the energy $E_0$ would be $P(E_0)=\abs{\bracket{0}{\psi}}^2=\abs{a}^2$. The probability of finding $E_1$ would be $P(E_1)=\abs{\bracket{1}{\psi}}^2=\abs{b}^2$. The expected value of the measurement is $\mean{H}= E_0\abs{a}^2+ E_1\abs{b}^2$. 

\vspace{0.25cm}
In the more general case of having a mixed state $\rho=\rho_{00} \proj{0} + \rho_{01} \op{ 0}{1} + \rho_{10} \op{1}{0} + \rho_{11} \proj{1}$ the probability of finding the ground state energy is $P(0)=\Tr \cor{ \proj{0} \rho }= \rho_{00}$, and the expected value of the energy would be $\mean{H}=\Tr \cor{H\rho}= E_0 \rho_{00} + E_1 \rho_{11}$.

\vspace{0.25cm} 
\end{minipage}
\label{minipage2}
}
\ec
\vspace{0.5cm}

\noindent
Another natural question to ask is how quantum systems evolve. The time-evolution of a pure state of a closed quantum system is given by the Schr\"odinger equation (see \cite{galindo_pascual_90}, Section 2.9). 

\vspace{0.5cm}
\begin{postulate} 
Time evolution of a pure state of a closed quantum system is given by the Schr\"odinger equation 

\be
\frac{d}{dt} \ket{\psi(t)} = -i\hbar H\ket{\psi(t)},
\label{eq:sch}
\ee
where $H$ is the {\it Hamiltonian} of the system and it is a Hermitian operator of the Hilbert space of the system state (from now on we avoid including Planck's constant by selecting the units such that $\hbar=1)$.
\label{post3}
\end{postulate}
\vspace{0.5cm}

\noindent
The Hamiltonian of a system is the operator corresponding to its energy, and it can be non-trivial to realise.  

Schr\"odinger equation can be formally solved in the following way. If at $t=0$ the state of a system is given by $\ket{\psi(0)}$ at time $t$ it will be 

\be
\ket{\psi(t)}=e^{-i Ht } \ket{\psi(0)}.
\ee
As $H$ is a Hermitian operator, the operator $U=e^{-i Ht }$ is  unitary. This gives us another way of phrasing Postulate \ref{post3}.

\vspace{0.5cm}
{\bf Postulate 3'}
{\it The evolution of a closed system is given by a unitary operator of the Hilbert space of the system }
\be
\ket{\psi(t)}=U \ket{\psi(0)},
\label{eq:evol}
\ee
{\it with} $U\in \cB\pare{\cH}$ {\it s.t.} $U U^{\dagger}=U^{\dagger}U=\id$.

\vspace{0.5cm}

\noindent
It is easy to prove that unitary operators preserve the norm of vectors and, therefore, transform pure states into pure states. As we did with the state of a system, it is reasonable to wonder if any unitary operator corresponds to the evolution of a real physical system. The answer is yes. 

\vspace{0.5cm}
\begin{lemma} 
All unitary evolutions of a state belonging to a finite Hilbert space can be constructed in several physical realisations like photons and cold atoms. 
\end{lemma}

\noindent
The proof of this lemma can be found at \cite{nielsen_00}.

The time evolution of a mixed state can be calculated just by combining Eqs. (\ref{eq:sch}) and (\ref{eq:dm}), giving the von-Neumann equation. 

\be
\dot{\rho} = - i \cor{H,\rho}\equiv \cL \rho,
\label{eq:vne}
\ee
where we have used the commutator $\cor{A,B}=AB-BA$, and $\cL$ is the so-called Liouvillian superoperator.

It is easy to prove that the Hamiltonian dynamics does not change the purity of a system

\be
\frac{d}{dt} \Tr\cor{\rho^2} = \Tr\cor{ \frac{d \rho^2}{dt} } = \Tr\cor{ 2\rho \dot{\rho} } = -2 i \Tr\cor{ \rho\pare{ H\rho -\rho H }  }=0,
\ee
where we have used the cyclic property of the trace. This result illustrates that the mixing rate of a state does not change due to the quantum evolution.

\bc
\framebox[15.5cm][l]{
\begin{minipage}[l]{15cm}
\vspace{0.25cm} 
{\bf Box 3. Time evolution of a two-level system.}
\vspace{0.25cm} 

The evolution of our isolated two-level system is described by its Hamiltonian
\be
H_{\text{free}}=E_0 \proj{0} + E_1 \proj{1},
\label{eq:atomham}
\ee
As the states $\ket{0}$ and $\ket{1}$ are Hamiltonian eigenstates if at $t=0$ the atom is at the excited state $\ket{\psi(0)}=\ket{1}$ after a time $t$ the state would be $\ket{\psi(t)}=e^{-iHt} \ket{1}=e^{-i E_1 t} \ket{1}$. 

\vspace{0.1cm}
As the system was already in an eigenvector of the Hamiltonian, its time-evolution consists only in adding a phase to the state, without changing its physical properties. (If an excited state does not change, why do atoms decay?) Without losing any generality we can fix the energy of the ground state as zero, obtaining 
\be
H_{\text{free}}= E \proj{1},
\label{eq:atomham2}
\ee
with $E\equiv E_1$. To make the model more interesting  we can include a driving that coherently switches between both states. The total Hamiltonian would be then

\be
H=E \proj{1} + \Omega \pare{\op{0}{1} +\op{1}{0}},
\ee
where $\Omega$ is the frequency of driving.  By using the von-Neumann equation (\ref{eq:vne}) we can calculate the populations $\pare{\rho_{00},\rho_{11}}$ as a function of time. The system is then driven between the states, and the populations present Rabi oscillations, as it is shown in Fig. \ref{figure1}. 

\bc
\includegraphics[scale=.8]{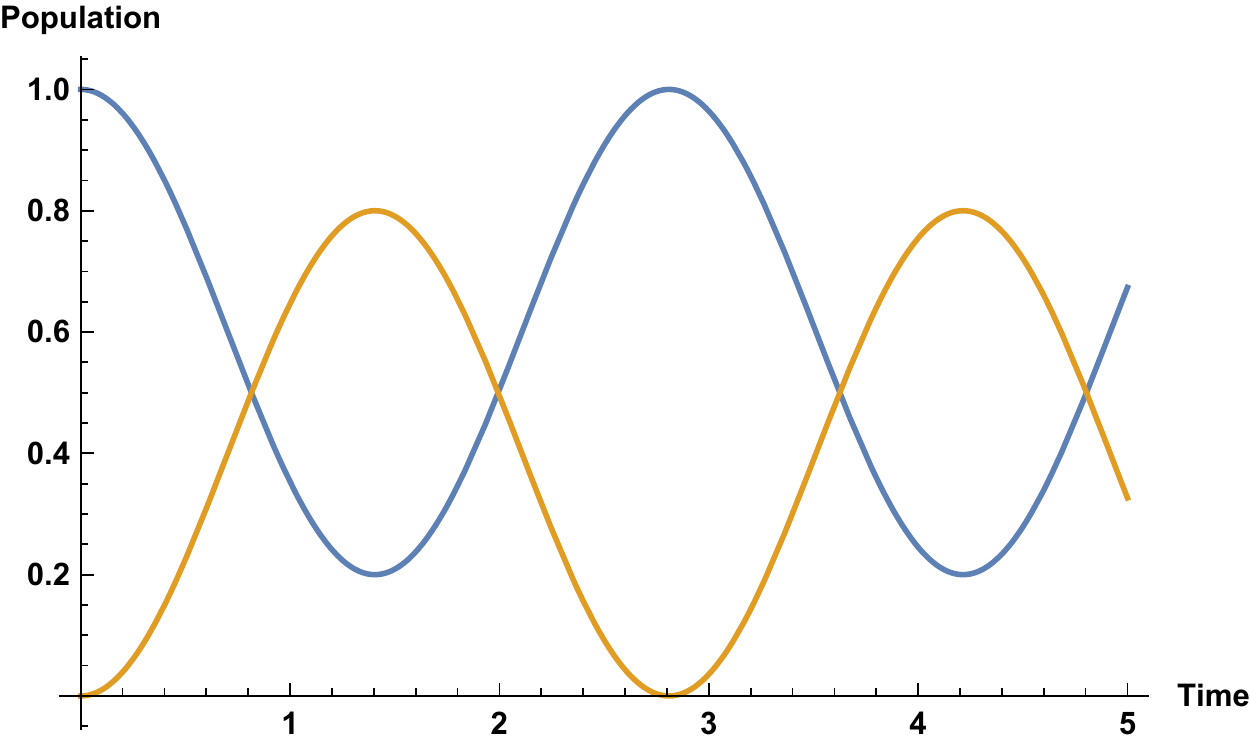}
\captionof{figure}{Population dynamics under a quantum dynamics (Parameters are $\Omega=1,\; E=1$). The blue line represents $\rho_{11}$ and the orange one $\rho_{00}$.}
\label{figure1}
\ec

\end{minipage}
}\ec
\vspace{0.5cm}

\noindent
Finally, as we are interested in composite quantum systems, we need to postulate how to work with them.

\vspace{0.5cm}
\begin{postulate} 
The state-space of a composite physical system, composed by $N$  subsystems, is the tensor product of the state space of each component
$\cH=\cH_1 \otimes \cH_2 \otimes  \cdots \otimes \cH_N$. The state of the composite physical system is given by a unit vector of $\cH$.  Moreover, if each subsystem belonging to $\cH_i$ is prepared in the state $\ket{\psi_i}$ the total state is given by $\ket{\psi}=\ket{\psi_1} \otimes \ket{\psi_2} \otimes \cdots \otimes\ket{\psi_N}$.
\end{postulate}
\vspace{0.5cm}

\noindent
The symbol $\otimes$ represents the tensor product of Hilbert spaces, vectors, and operators. If we have a composited mixed state where each component is prepared in the state $\rho_i$ the total state is given by $\rho=\rho_1 \otimes \rho_2 \otimes \cdots \otimes\rho_N$. 

States that can be expressed in the simple form  $\ket{\psi}=\ket{\psi_1} \otimes \ket{\psi_2}$, in any specific basis, are very particular and they are called {\it separable states} (For this discussion, we use a bipartite system as an example. The extension to a general multipartite system is straightforward.) . In general, any arbitrary state should be described as $\ket{\psi}=\sum_{i,j} \ket{\psi_i}  \otimes \ket{\psi_j}$ (or $\rho=\sum_{i,j} \rho_i \otimes \rho_j$  for mixed states). Non-separable states are called {\it entangled states}. 

Now that  we know how to compose systems, but we can be interested in going the other way around. If we have a system belonging to a bipartite Hilbert space in the form   $\cH=\cH_a \otimes \cH_b$ we can be interested in studying some properties of the subsystem corresponding to one of the subspaces. To do so, we define the {\it reduced density matrix}. If the state of our system is described by a density matrix $\rho$ the reduced density operator of the subsystem $a$ is defined by the operator 

\be
\rho_{a} \equiv \Tr_{b} \cor{\rho},
\ee
were $\Tr_b$ is the partial trace over subspace $b$ and it is defined as \cite{nielsen_00}

\be
\Tr_b \cor{ \sum_{i,j,k,l} \op{a_i}{a_j} \otimes \op{b_k}{b_l} } \equiv\sum_{i,j} \op{a_i}{a_j} \Tr \cor{ \sum_{k,l} \op{b_k}{b_l}}.
\ee
The concepts of reduced density matrix and partial trace are essential in the study of open quantum systems. If we want to calculate the equation of motions of a system affected by an environment, we should trace out this environment and deal only with the reduced density matrix of the system. This is the main idea of the theory of open quantum systems.


\newpage
\bc
\framebox[15.5cm][l]{
\begin{minipage}[l]{15cm}
\vspace{0.25cm} 

{\bf Box 4. Two two-level atoms}
\vspace{0.25cm} 

If we have two two-level systems, the total Hilbert space is given by $\cH=\cH_2\otimes \cH_2$. A basis of this Hilbert space would be given by the set $\left\{ \ket{00}\equiv  \ket{0}_1 \otimes\ket{0}_2,\; \ket{01}\equiv  \ket{0}_1 \otimes\ket{1}_2,\; \ket{10}\equiv  \ket{1}_1 \otimes\ket{0}_2,\;\ket{11}\equiv  \ket{1}_1 \otimes\ket{1}_2 \right\}$. If both systems are in their ground state, we can describe the total state by the separable vector 

\be
\ket{\psi}_G=\ket{00}.
\ee 
A more complex, but still separable, state can be formed if both systems are in superposition. 

\ben
\ket{\psi}_S&=&\frac{1}{\sqrt{2}} \left( \ket{0}_1 +\ket{1}_1 \right) \otimes \frac{1}{	\sqrt{2}} \left( \ket{0}_2 +\ket{1}_2 \right) \nonumber\\
&=& \frac{1}{2} \left( \ket{00} + \ket{10} + \ket{01} + \ket{11}  \right)
\een

An entangled state would be 

\be
\ket{\psi}_E=\frac{1}{	\sqrt{2}} \left( \ket{00} +\ket{11} \right).
\ee
This state cannot be separated into a direct product of each subsystem. If we want to obtain a reduced description of subsystem $1$ (or $2$) we have to use the partial trace. To do so, we need first to calculate the density matrix corresponding to the pure state $\ket{\psi}_E$. 

\be
\rho_E=\ket{\psi}\bra{\psi}_E = \frac{1}{2} \left( \proj{00} + \op{00}{11} + \op{11}{00} + \proj{11} \right).
\ee
We can now calculate the reduced density matrix of the subsystem $1$ by using the partial trace. 

\be
\rho_E^{(1)}= \bra{0}_2  \rho_E \ket{0}_2 +  \bra{1}_2  \rho_E \ket{1}_2  = \frac{1}{2} \left( \proj{00}_1 + \proj{11}_2 \right).
\ee 
From this reduced density matrix, we can calculate all the measurement statistics of subsystem $1$. 

\vspace{0.25cm}

\end{minipage}
}
\ec

\vspace{0.5cm}

\newpage
\section{The Fock-Liouville Hilbert space. The Liouville superoperator}
\label{sec:fl}

In this section, we revise a useful framework for both analytical and numerical calculations. It is clear that some linear combinations of density matrices are valid density matrices (as long as they preserve positivity and trace $1$). Because of that, we can create a Hilbert space of density matrices just by defining a scalar product. This is clear for finite systems because in this case scalar space and Hilbert space are the same things. It also happens to be true for infinite spaces. This allows us to define a linear space of matrices, converting the matrices effectively into vectors ($\rho\to\kket{\rho}$).  This is called Fock-Liouville space (FLS). The usual definition of the scalar product of matrices $\phi$ and $\rho$ is defined as  $\bbracket{\phi}{\rho}\equiv\Tr\cor{\phi^\dagger\rho}$. The Liouville super-operator from Eq. (\ref{eq:vne}) is now an operator acting on the Hilbert space of density matrices. The main utility of the FLS is to allow the matrix representation of the evolution operator. 

\vspace{0.5cm}
\bc
\framebox[15.5cm][l]{
\begin{minipage}[l]{15cm}
\vspace{0.25cm} 

{\bf  Box 5. Time evolution of a two-level system.}
\vspace{0.25cm}

The density matrix of our system (\ref{eq:denmat}) can be expressed in the FLS as 

\be
\kket{\rho}=
\begin{pmatrix}
\rho_{00} \\
\rho_{01} \\
\rho_{10} \\
\rho_{11} 
\end{pmatrix}.
\ee
\vspace{0.25cm}

The time evolution of a mixed state is given by the von-Neumann equation (\ref{eq:vne}). The Liouvillian superoperator can now be expressed as a matrix

\be
\tilde{\cL}= 
\left(
\begin{array}{cccc}
0  &  i \Omega  & -i\Omega & 0 \\
i\Omega &  i E  &0 & -i\Omega \\
-i\Omega   & 0  & -iE & i\Omega \\
0 & -i\Omega  &  i\Omega & 0 
\end{array}
\right),
\ee
where each row is calculated just by observing the output of the operation $-i \cor{H,\rho}$ in the computational basis of the density matrices space. The time evolution of the system now corresponds to the matrix equation $\frac{d \kket{\rho}}{dt}=\tilde{\cL} \kket{\rho}$, that in matrix notation  would be

\be
\begin{pmatrix}
\dot{\rho}_{00} \\
\dot{\rho}_{01} \\
\dot{\rho}_{10} \\
\dot{\rho}_{11} 
\end{pmatrix}
=
\left(
\begin{array}{cccc}
0  &  i \Omega  & -i\Omega & 0 \\
i\Omega &  i E  &0 & -i\Omega \\
-i\Omega   & 0  & -i E & i\Omega \\
0 & -i\Omega  &  i\Omega & 0 
\end{array}
\right)
\begin{pmatrix}
\rho_{00} \\
\rho_{01} \\
\rho_{10} \\
\rho_{11} 
\end{pmatrix}
\ee

\vspace{0.25cm}
\label{minipage4}
\end{minipage}
}
\ec
\vspace{0.5cm}


\newpage
\section{CPT-maps and the Lindblad Master Equation.}
\label{cpt}

\subsection{Completely positive maps}

The  problem we want to study is to find the most general Markovian transformation set between density matrices. Until now, we have seen that quantum systems can evolve in two way, by a coherent evolution given (Postulate \ref{post3}) and by collapsing after a measurement (Postulate \ref{post4}). Many efforts have been made to unify these two ways of evolving \cite{schlosshauer_07}, without giving a definite answer so far. It is reasonable to ask what is the most general transformation that can be performed in a quantum system, and what is the dynamical equation that describes this transformation. 

We are looking for maps that transform density matrices into density matrices. We define $\rho(\cH)$ as the space of all density matrices in the Hilbert space $\cH$. Therefore, we are looking for a map of this space onto itself, $\cV:\rho(\cH)\to\rho(\cH)$. To ensure that the output of the map is a density matrix this should fulfil the following properties

\begin{itemize}
\item Trace preserving. $\Tr\cor{\cV A}=\Tr\cor{A},$ $\forall A\in O(\cH)$.
\item Completely positive (see below). 
\end{itemize}

Any map that fulfils these two properties is called a {\it completely positive and trace-preserving map (CPT-maps)}. The first property is quite apparent, and it does not require more thinking. The second one is a little more complicated, and it requires an intermediate definition. 

\vspace{0.5cm}
\begin{definition}
A map $\cV$ is positive iff $\forall A\in B(\cH)$ s.t.  $A \ge 0 \Rightarrow \cV A \ge 0$.
\end{definition}
\vspace{0.5cm}

\noindent
This definition is based in the idea that, as density matrices are positive, any physical map should transform positive matrices into positive matrices. One could naively think that this condition must be sufficient to guarantee the physical validity of a map. It is not. As we know, there exist composite systems, and our density matrix could be the partial trace of a more complicated state. Because of that, we need to impose a more general condition. 

\vspace{0.5cm}
\begin{definition}
A map $\cV$ is completely positive iff $\forall n\in \mathbb{N}$, $\cV\otimes \id_n$ is positive.
\end{definition}
\vspace{0.5cm}

\noindent
To prove that not all positive maps are completely positive, we need a counterexample. A canonical example of an operation that is positive but fails to be completely positive is the matrix transposition. If we have a Bell state in the form $\ket{\psi_B}=\frac{1}{\sqrt{2}} \pare{\ket{01}+\ket{10}}$ its density matrix can be expressed as 

\be
\rho_B=\frac{1}{2} \pare{\op{0}{0} \otimes \op{1}{1} + \op{1}{1} \otimes \op{0}{0} + \op{0}{1} \otimes\op{1}{0} + \op{1}{0} \otimes\op{0}{1}},
\ee
with a matrix representation 

\ben
\rho_B= \frac{1}{2} \left\{ 
\left(
\begin{array}{cc}
1 & 0 \\
0 & 0 \\
\end{array}
\right) 
\otimes 
\left(
\begin{array}{cc}
0 & 0 \\
0 & 1 \\
\end{array}
\right) 
+
\left(
\begin{array}{cc}
0 & 0 \\
0 & 1 \\
\end{array}
\right) 
\otimes 
\left(
\begin{array}{cc}
1 & 0 \\
0 & 0 \\
\end{array}
\right) 
\right.
\nonumber\\
\left.
\otimes 
\left(
\begin{array}{cc}
0 & 0 \\
1 & 0 \\
\end{array}
\right) 
\otimes 
\left(
\begin{array}{cc}
0 & 1 \\
0 & 0 \\
\end{array}
\right) 
+
\left(
\begin{array}{cc}
0 & 1 \\
0 &  0\\
\end{array}
\right) 
\otimes 
\left(
\begin{array}{cc}
0 & 0 \\
1 & 0 \\
\end{array}
\right) 
\right\}.
\een
A little algebra shows that the full form of this matrix is 

\be
\rho_B=\left(
\begin{array}{cccc}
0 & 0 & 0 & 0\\
0 & 1 & 1 & 0\\
0 & 1 & 1 & 0\\
0 & 0 & 0 & 0
\end{array}
\right),
\ee
and it is positive. 

It is easy to check that the transformation $\id\otimes T_2 $, meaning that we transpose the matrix of the second subsystem leads to a non-positive matrix 

\ben
\pare{ \id\otimes T_2 } \rho_B= \frac{1}{2} \left\{ 
\left(
\begin{array}{cc}
1 & 0 \\
0 & 0 \\
\end{array}
\right) 
\otimes 
\left(
\begin{array}{cc}
0 & 1 \\
0 & 0 \\
\end{array}
\right) 
+
\left(
\begin{array}{cc}
0 & 0 \\
0 & 1 \\
\end{array}
\right) 
\otimes 
\left(
\begin{array}{cc}
0 & 0 \\
1 & 0 \\
\end{array}
\right) 
\right.
\nonumber\\
\left.
\otimes 
\left(
\begin{array}{cc}
0 & 0 \\
1 & 0 \\
\end{array}
\right) 
\otimes 
\left(
\begin{array}{cc}
0 & 0 \\
1 & 0 \\
\end{array}
\right) 
+
\left(
\begin{array}{cc}
0 & 0 \\
0 & 1 \\
\end{array}
\right) 
\otimes 
\left(
\begin{array}{cc}
0 & 1 \\
0 & 0 \\
\end{array}
\right) 
\right\}.
\een
The total matrix is 

\be
\pare{ \id\otimes T_2 } \rho_B=
\left(
\begin{array}{cccc}
0 & 0 & 0 & 1\\
0 & 1 & 0 & 0\\
0 & 0 & 1 & 0\\
1 & 0 & 0 & 0
\end{array}
\right),
\ee
with $-1$ as an eigenvalue. This example illustrates how the non-separability of quantum mechanics restrict the operations we can perform in a subsystem. By imposing this two conditions, we can derive a unique master equation as the generator of any possible Markovian CPT-map. 

\subsection{Derivation of the Lindblad Equation from microscopic dynamics}

\begin{figure}
\begin{center}
\includegraphics[scale=0.2]{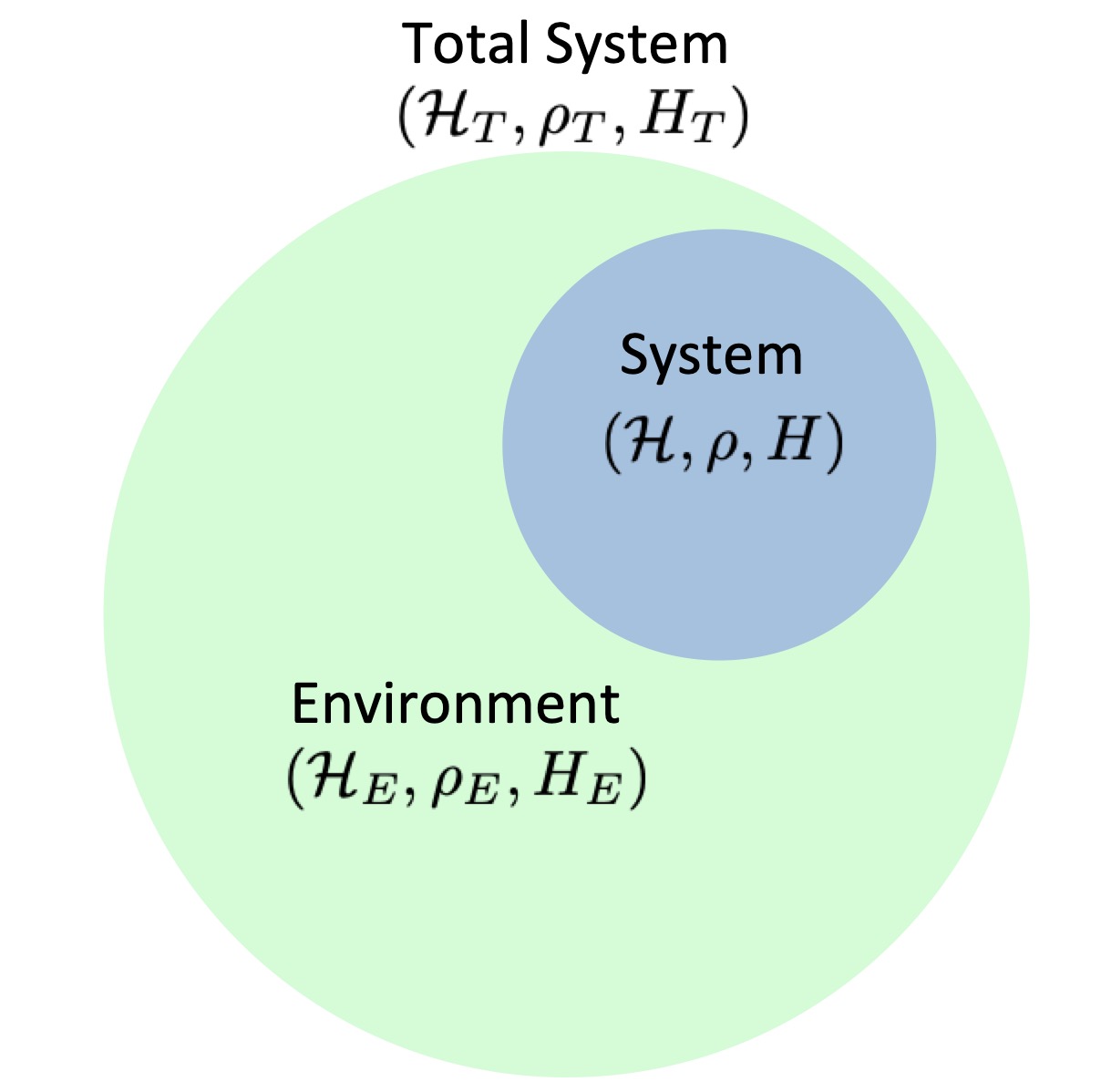}
\end{center}
\caption{A total system (belonging to a Hilbert space $\cH_T$, with states described by density matrices $\rho_T$, and with dynamics determined by a Hamiltonian $H_T$) divided into the system of interest, `System', and the environment.   }
\label{fig:fig01}
\end{figure}

The most common derivation of the Lindblad master equation is based on Open Quantum Theory. The Lindblad equation is then an effective motion equation for a subsystem that belongs to a more complicated system. This derivation can be found in several textbooks like Breuer and Petruccione's \cite{breuer_02} as well as Gardiner and Zoller's \cite{gardiner_00}. Here, we follow the derivation presented in Ref. \cite{manzano:av18}. Our initial point is displayed in Figure \ref{fig:fig01}. A total system belonging to a Hilbert space $\cH_T$ is divided into our system of interest, belonging to a Hilbert space $\cH$, and the environment living in $\cH_E$. 

The evolution of the total system is given by the von Neumann equation (\ref{eq:vne}). 

\be
\dot{\rho_T}(t)=-i \left[ H_T,\rho_T(t) \right].
\ee
As we are interested in the dynamics of the system, without the environment, we trace over the environment degrees of freedom to obtain the reduced density matrix of the system $\rho(t)=\Tr_E[\rho_T]$. To separate the effect of the total hamiltonian in the system and the environment we divide it in the form $H_T=H_S \otimes \id_E + \id_S \otimes H_E + \alpha H_I $, with $H\in\cH$, $H_E\in \cH_E$, and $H_I\in \cH_T$, and being $\alpha$ a measure of the strength of the system-environment interaction. Therefore, we have a part acting on the system, a part acting on the environment, and the interaction term. Without losing any generality, the interaction term can be decomposed in the following way 

\be
H_I =  \sum_i S_i \otimes E_i,
\label{eq:hint}
\ee
with $S_i\in B(\cH$) and  $E_i\in B(\cH_E)$\footnote{From now on we will not writethe identity operators of the Hamiltonian parts explicitly  when they can be inferred from the context.}.

To better describe the dynamics of the system, it is useful to work in the interaction picture (see Ref. \cite{galindo_pascual_90} for a detailed explanation about Schr\"odinger, Heisenberg, and interaction pictures). In the interaction picture,  density matrices evolve with time due to the interaction Hamiltonian, while operators evolve with the system and environment Hamiltonian. An arbitrary operator $O\in \cB(\cH_T)$ is represented in this picture by the time-dependent operator $\hat{O}(t)$, and its time evolution is

\be
\hat{O}(t) = e^{i(H+H_E)t} \,O\, e^{-i(H+H_E)t}.
\ee
The time evolution of the total density matrix is given in this picture by 

\be
\frac{d \hat{\rho}_T(t)}{dt} = -i \alpha \cor{\hat{H}_I(t),\hat{\rho}_T(t)}.
\label{eq:vnint}
\ee
This equation can be easily integrated to give 

\be
\hat{\rho}_T(t) = \hat{\rho}_T(0) -i \alpha \int_0^t ds \cor{\hat{H}_I(s),\hat{\rho}_T(s)}.
\label{eq:integint}
\ee
By this formula, we can obtain the exact solution, but it still has the complication of calculating an integral in the total Hilbert space. It is also troublesome the fact that the state $\tilde{\rho}(t)$ depends on the integration of the density matrix in all previous time. To avoid that we can introduce Eq. (\ref{eq:integint}) into Eq. (\ref{eq:vnint}) giving 

\be
\frac{d \hat{\rho}_T(t)}{dt} = -i \alpha \cor{\hat{H}_I(t),\hat{\rho}_T(0)} -\alpha^2 \int_0^t ds \cor{ \hat{H}_I(t),\cor{\hat{H}_I(s),\hat{\rho}_T(s)} }.
\ee
By applying this method one more time we obtain 

\be
\frac{d \hat{\rho}_T(t)}{dt} = -i \alpha \cor{\hat{H}_I(t),\hat{\rho}_T(0)} -\alpha^2 \int_0^t  ds \cor{ \hat{H}_I(t),\cor{\hat{H}_I(s),\hat{\rho}_T(t)} } + O(\alpha^3).
\label{eq:orderthree}
\ee
After this substitution, the integration of the previous states of the system is included only in the terms that are $O(\alpha^3)$ or higher. At this moment, we perform our first approximation by considering that the strength of the interaction between the system and the environment is small. Therefore, we can avoid high-orders in Eq. (\ref{eq:orderthree}). Under this approximation we have

\be
\frac{d \hat{\rho}_T(t)}{dt} = -i \alpha \cor{\hat{H}_I(t),\hat{\rho}_T(0)} -\alpha^2 \int_0^t  ds \cor{ \hat{H}_I(t),\cor{\hat{H}_I(s),\hat{\rho}_T(t)} }.
\label{eq:exp}
\ee
We are interested in finding an equation of motion for $\rho$, so we trace over the environment degrees of freedom 

\be
\frac{d \hat{\rho}(t)}{dt}= \Tr_E \left[\frac{d \hat{\rho}_T(t)}{dt} \right] = -i \alpha \Tr_E\cor{\hat{H}_I(t),\hat{\rho}_T(0)} -\alpha^2 \int_0^t  ds \Tr_E\cor{ \hat{H}_I(t),\cor{\hat{H}_I(s),\hat{\rho}_T(t)} }.
\label{eq:exp2}
\ee
This is not a closed time-evolution equation for $\hat{\rho}(t)$, because the time derivative still depends on the full density matrix $\hat{\rho}_T(t)$. To proceed, we need to make two more assumptions. First, we assume that $t=0$ the system and the environment have a separable state in the form $\rho_T(0)=\rho(0) \otimes \rho_E(0)$. This means that there are not correlations between the system and the environment. This may be the case if the system and the environment have not interacted at previous times or if the correlations between them are short-lived. Second, we assume that the initial state of the environment is thermal, meaning that it is described by a density matrix in the form $\rho_E(0)=\exp\left( -H_E/T \right)/\Tr[\exp\left( -H_E/T \right)]$, being $T$ the temperature and taking the Boltzmann constant as $k_B=1$. By using these assumptions, and the expansion of $H_I$ (\ref{eq:hint}), we can calculate an expression for  the first element of the r.h.s of Eq. (\ref{eq:exp2}). 

\be
\Tr_E\cor{\hat{H}_I(t),\hat{\rho}_T(0)} = \sum_i \left( \hat{S}_i(t) \hat{\rho}(0) \Tr_E \left[ \hat{E}_i(t) \hat{\rho}_E(0)  \right]  -     
\hat{\rho}(0) \hat{S}_i(t)   \Tr_E \left[ \hat{\rho}_E(0) \hat{E}_i(t)   \right]  \right).
\label{eq:zero}
\ee
To calculate the explicit value of this term, we may use that $\left< E_i\right>=\Tr[E_i \rho_E(0)]=0$ for all values of $i$. This looks like a strong assumption, but it is not. If our total Hamiltonian does not fulfil it, we can always rewrite it as 
$H_T=\left( H+ \alpha \sum_i  \left< E_i\right>  S_i\right) + H_E + \alpha H_i'$, with $H'_i= \sum_i S_i \otimes (E_i- \left< E_i\right>)$. It is clear that now $\left< E'_i \right>=0$, with $E'_i=E_i- \left< E_i\right>$, and the system Hamiltonian is changed just by the addition of an energy shift that does no affect  the system dynamics. Because of that, we can assume that $\left< E_i\right>=0$ for all $i$. Using the cyclic property of the trace, it is easy to prove that the term of Eq. (\ref{eq:zero}) is equal to zero, and the equation of motion (\ref{eq:exp2}) reduces to 

\be
\dot{\hat{\rho}}(t)=  -\alpha^2 \int_0^t  ds \Tr_E\cor{ \hat{H}_I(t),\cor{\hat{H}_I(s),\hat{\rho}_T(t)} }.
\label{eq:exp3}
\ee
This equation still includes the entire state of the system and environment. To unravel the system from the environment, we have to make a more restrictive assumption. As we are working in the weak coupling regime, we may suppose that the system and the environment are non-correlated during all the time evolution. Of course, this is only an approximation. Due to the interaction Hamiltonian, some correlations between system and environment are expected to appear. On the other hand, we may assume that the timescales of correlation ($\tau_\text{corr}$) and relaxation of the environment ($\tau_\text{rel}$) are much smaller than the typical system timescale ($\tau_\text{sys}$), as the coupling strength is very small ($\alpha<<$). Therefore, under this strong assumption, we can assume that the environment state is always thermal and is decoupled from the system state, $\hat{\rho}_T(t)=\hat{\rho}(t) \otimes \hat{\rho}_E(0)$. Eq. (\ref{eq:exp3}) then transforms into 

\be
\dot{\hat{\rho}}(t)=  -\alpha^2 \int_0^t  ds \Tr_E\cor{ \hat{H}_I(t),\cor{\hat{H}_I(s),\hat{\rho}(t) \otimes \hat{\rho}_E(0)}  }.
\ee
The equation of motion is now independent for the system and local in time. It is still non-Markovian, as it depends on the initial state preparation of the system. We can obtain a Markovian equation by realising that the kernel in the integration and that we can extend the upper limit of the integration to infinity with no real change in the outcome. By doing so, and by changing the  integral variable to $s\rightarrow t-s$, we obtain the famous {\it Redfield equation} \cite{redfield:IBM57}.

\be
\dot{\hat{\rho}}(t)=  -\alpha^2 \int_0^{\infty}  ds \Tr_E\cor{ \hat{H}_I(t),\cor{\hat{H}_I(s-t),\hat{\rho}(t) \otimes \hat{\rho}_E(0)}  }.
\label{eq:redfield}
\ee
It is known that this equation does not warrant the positivity of the map, and it sometimes gives rise to density matrices that are non-positive. To ensure complete positivity, we need to perform one further approximation, the {\it rotating wave} approximation. To do so, we need to use the spectrum of the superoperator $\tilde{H}A\equiv \cor{H,A}$,  $\forall A\in \cB(\cH)$. The eigenvectors of this superoperator form a complete basis of space $\cB(\cH)$ and, therefore, we can expand the system-environment operators from Eq. (\ref{eq:hint}) in this basis

\be
S_i = \sum_{\omega} S_i(\omega), 
\label{eq:expeigen}
\ee
where the operators $S_i(\omega)$ fulfils 

\be
\cor{H,S_i(\omega)}= -\omega S_i(\omega),
\ee
being $\omega$ the eigenvalues of $\tilde{H}$. It is easy to take also the Hermitian conjugated 

\be
\cor{H,S_i^{\dagger}(\omega)}= \omega S_i^{\dagger}(\omega).
\ee
To apply this decomposition, we need to change back to the Schr\"odinger picture for the term of the interaction Hamiltonian acting on the system's Hilbert space. This is done by the expression $\hat{S}_k= e^{it H} S_k e^{-it H}$. By using the eigen-expansion (\ref{eq:expeigen}) we arrive to 

\be
\tilde{H}_i(t) = \sum_{k,\omega} e^{-i\omega t} S_k(\omega) \otimes \tilde{E}_k (t) = \sum_{k,\omega} e^{i\omega t} S_k^{\dagger}(\omega) \otimes \tilde{E}_k^{\dagger} (t).
\ee
To combine this decomposition with Redfield equation (\ref{eq:redfield}), we first may expand the commutators. 

\ben
\hspace{-2cm}
\dot{\hat{\rho}}(t)= -\alpha^2 \Tr\left[ \int_0^\infty ds\, \hat{H}_I (t) \hat{H}_I (t-s) \hat{\rho} (t) \otimes \hat{\rho}_E(0)  
- \int_0^\infty ds\, \hat{H}_I (t) \hat{\rho} (t) \otimes \hat{\rho}_E(0) \hat{H}_I (t-s)    \right. \nonumber \\
\hspace{-2cm}
\left. - \int_0^\infty ds\, \hat{H}_I (t-s) \hat{\rho} (t) \otimes \hat{\rho}_E(0) \hat{H}_I (t) 
+ \int_0^\infty ds\, \hat{\rho} (t) \otimes \hat{\rho}_E(0) \hat{H}_I (t-s) \hat{H}_I (t)   
\right].
\een
We now apply the eigenvalue decomposition in terms of $S_k(\omega)$ for $\hat{H}_I(t-s)$ and in terms of $S_k^{\dagger}(\omega')$ for $\hat{H}_I(t)$. By using the permutation property of the trace and the fact that $\cor{H_E,\rho_E(0)}=0$, and after some non-trivial algebra we obtain 

\be
\dot{\hat{\rho}}(t) =\sum_{\substack{\omega,\omega'\\ k,l }} \left( e^{i (\omega'-\omega)t }\, \Gamma_{kl} (\omega) \cor{S_l(\omega)\hat{\rho} (t),  S_k^\dagger(\omega') }   +
 e^{i (\omega-\omega')t } \, \Gamma_{lk}^* (\omega') \cor{S_l(\omega),  \hat{\rho} (t)  S_k^\dagger(\omega') }  \right),
\label{eq:rwae}
\ee
where the effect of the environment has been absorbed into the factors 

\be
\Gamma_{kl} (\omega) \equiv  \int_0^{\infty} ds\, e^{i\omega s} \Tr_E \left[ \tilde{E}_k^\dagger (t) \tilde{E}_l (t-s) \rho_E(0)  \right],
\ee
where we are writing the environment operators of the interaction Hamiltonian in the interaction picture ($\hat{E}_l(t)=e^{iH_Et}  E_l e^{-iH_Et} $). At this point, we can already perform the rotating wave approximation. By considering the time-dependency on Eq. (\ref{eq:rwae}), we conclude that the terms with $\left| \omega-\omega' \right|>>\alpha^2$ will oscillate much faster than the typical timescale of the system evolution. Therefore, they do not contribute to the evolution of the system.  In the low-coupling regime $(\alpha\rightarrow 0)$ we can consider that only the resonant terms, $\omega=\omega'$, contribute to the dynamics and remove all the others. By applying this approximation to Eq. (\ref{eq:rwae}) reduces to 

\be
\dot{\hat{\rho}}(t) =\sum_{\substack{\omega\\ k,l }} \left( \Gamma_{kl} (\omega) \cor{S_l(\omega)\hat{\rho} (t),  S_k^\dagger(\omega) }   +
 \Gamma_{lk}^* (\omega) \cor{S_l(\omega),  \hat{\rho} (t)  S_k^\dagger(\omega) }  \right).
\ee
To divide the dynamics into Hamiltonian and non-Hamiltonian we now decompose the operators $\Gamma_{kl}$ into Hermitian and non-Hermitian parts, $\Gamma_{kl}(\omega) = \frac{1}{2} 	 \gamma_{kl}(\omega) +  i\pi_{kl} $, with

\ben
\pi_{kl}(\omega) \equiv \frac{-i}{2} \left(\Gamma_{kl} (\omega) - \Gamma_{kl}^*(\omega)   \right)  \nonumber\\
\gamma_{kl}(\omega) \equiv  \Gamma_{kl} (\omega) + \Gamma_{kl}^*(\omega) = \int_{-\infty}^{\infty} ds e^{i\omega s}
\Tr\left[ \hat{E}_k^\dagger (s) E_l \hat{\rho}_E(0)\right].
\een 
By these definitions we can separate the Hermitian and non-Hermitian parts of the dynamics and we can transform back to the Schr\"odinger picture

\be
\dot{\rho}(t) = -i \left[ H+H_{Ls} ,\rho(t) \right] + 
\sum_{\substack{\omega\\ k,l }} \gamma_{kl} (\omega) \left(  S_l (\omega) \rho(t) S_k^\dagger (\omega) - 
\frac{1}{2}  \left\{ S_k^\dagger S_l (\omega) , \rho(t) \right\} \right).
\label{eq:me1}
\ee
The Hamiltonian dynamics now is influenced by a term $H_{Ls} = \sum_{\omega,k,l}  \pi_{kl} (\omega)   S_k^\dagger (\omega)S_l (\omega)$. This is usually called a {\it Lamb shift} Hamiltonian and its role is to renormalize the system energy levels due to the interaction with the environment. Eq. (\ref{eq:me1}) is the first version of the Markovian Master Equation, but it is not in the Lindblad form yet. 

It can be easily proved that the matrix formed by the coefficients $\gamma_{kl}(\omega)$ is positive as they are the Fourier's transform of a positive function $\left(\Tr\left[ \hat{E}_k^\dagger (s) E_l \hat{\rho}_E(0)\right]\right)$. Therefore, this matrix can be diagonalised. This means that we can find a unitary operator, $O$, s.t. 

\be
O\gamma(\omega) O^\dagger=
\left(\begin{array}{cccc}
d_1(\omega) & 0 & \cdots & 0\\
0 & d_2(\omega) & \cdots & 0 \\
\vdots & \vdots & \ddots & 0\\
0 & 0 & \cdots & d_N(\omega)
\end{array}\right).
\ee
We can now write the master equation in a diagonal form 

\be
\dot{\rho}(t) = -i \left[ H+H_{Ls} ,\rho(t) \right] + 
\sum_{i,\omega}  \left(  L_i (\omega) \rho(t) L_i^\dagger (\omega) - 
\frac{1}{2}  \left\{ L_i^\dagger L_i (\omega) , \rho(t) \right\} \right)\equiv \cL \rho(t).
\label{eq:me2}
\ee
This is the celebrated Lindblad (or Lindblad-Gorini-Kossakowski-Sudarshan) Master Equation. In the simplest case, there will be only one relevant frequency $\omega$, and the equation can be further simplified to 

\be
\dot{\rho}(t) = -i \left[ H+H_{Ls} ,\rho(t) \right] + 
\sum_{i}  \left(  L_i \rho(t) L_i^\dagger - 
\frac{1}{2}  \left\{ L_i^\dagger L_i, \rho(t) \right\} \right)\equiv \cL \rho(t).
\label{eq:me3}
\ee
The operators $L_i$ are usually referred to as {\it jump operators}.

\subsection{Derivation of the Lindblad Equation as a CPT generator}

The second way of deriving Lindblad equation comes from the following question: What is the most general (Markovian) way of mapping density matrix onto density matrices? This is usually the approach from quantum information researchers that look for general transformations of quantum systems. We analyse this problem following mainly Ref. \cite{wilde_17}. 

To start, we need to know what is the form of a general CPT-map.

\vspace{0.5cm}
\begin{lemma}
Any map $\cV:B\pare{\cH}\to B\pare{\cH}$ that can be written in the form $\cV\rho=V^\dagger \rho V^{\phantom{\dagger}}$ with $V\in B\pare{\cH}$ is positive.
\end{lemma}
\vspace{0.5cm}
\noindent
The proof of the lemma requires a little algebra and a known property of normal matrices 

\vspace{0.5cm}
\noindent
\textbf{Proof.}

\noindent
If $\rho\ge0 \Rightarrow \rho=A^\dagger A^{\phantom{\dagger}}$, with $A\in B(\cH)$. Therefore, $\cV\rho=V^\dagger\rho V^{\phantom{\dagger}}\Rightarrow \bra{\psi} V^\dagger\rho V\ket{\psi} = \bra{\psi} V^\dagger A^{\dagger}A V\ket{\psi}=\norm{ AV\ket{\psi}}\ge 0$.  Therefore, if $\rho$ is positive, the output of the map is also positive.

\vspace{0.25 cm}
\noindent
\textbf{End of the proof.}
\vspace{0.5cm}

\noindent
This is a sufficient condition for the positivity of a map, but it is not necessary. It could happen that there are maps that cannot be written in this form, but they are still positive. To go further, we need a more general condition, and this comes in the form of the next theorem.

\vspace{0.5cm}
\begin{theorem}
Choi's Theorem. 

\noindent
A linear map $\cV:B(\cH)\to B(\cH)$ is completely positive iff it can be expressed as 

\be
\cV\rho=\sum_i V_i^\dagger \rho V^{\phantom{\dagger}}_i
\label{eq:choimap}
\ee
with $V_i\in B(\cH)$. 
 \end{theorem}
\vspace{0.5cm}

\noindent
The proof of this theorem requires some algebra. 

\vspace{0.5cm}
\noindent
\textbf{Proof}

\noindent
The `if' implication is a trivial consequence of the previous lemma. To prove the converse, we need to extend the dimension of our system by the use of an auxiliary system. If $d$ is the dimension of the Hilbert space of pure states, $\cH$, we define a new Hilbert space of the same dimension $\cH_A$. 

We define a maximally entangled pure state in the bipartition $\cH_A \otimes \cH$ in the way 

\be
\ket{\Gamma}\equiv \sum_{i=0}^{d-1} \ket{i}_A \otimes \ket{i},
\label{eq:gamma}
\ee
being $\key{\ket{i}}$ and ${\key{\ket{i}_A}}$ arbitrary orthonormal bases for  $\cH$ and $\cH_A$. 

We can extend the action of our original map $\cV$, that acts on $\cB(\cH)$ to our extended Hilbert space by defining the map $\cV_2:\cB( \cH_A) \otimes \cB(\cH) \to \cB( \cH_A) \otimes \cB(\cH)$ as 

\be
\cV_2\equiv \id_{\cB( \cH_A)} \otimes \cV.
\ee
Note that the idea behind this map is to leave the auxiliary subsystem invariant while applying the original map to the original system. This map is positive because $\cV$ is completely positive.  This may appear trivial, but as it has been explained before complete positivity is a more restrictive property than positivity, and we are looking for a condition to ensure complete positivity. 

We can now apply the extended map to the density matrix corresponding to the maximally entangled state (\ref{eq:gamma}), obtaining

\be
\cV_2 \proj{\Gamma} = \sum_{i,j=0}^{d-1} \op{i}{j}  \otimes \cV \op{i}{j}.
\label{eq:map2}
\ee
Now we can use the maximal entanglement of the state $\ket{\Gamma}$ to relate the original map $\cV$ and the action $\cV_2 \proj{\Gamma}$ by taking the matrix elements with respect to $\cH_A$. 

\be
\cV\op{i}{j} = \bra{i}_A \pare{ \cV_2\op{\Gamma}{\Gamma} }\ket{j}_A.
\label{eq:elementsv}
\ee
To relate this operation to the action of the map to an arbitrary vector $\ket{\psi}\in \cH_A \otimes  \cH$, we can expand it in this basis as

\be
\ket{\psi} = \sum_{i=0}^{d-1} \sum_{j=0}^{d-1} \alpha_{ij} \ket{i}_A \otimes \ket{j}.
\ee
We can also define an operator $V_{\ket{\psi}} \in \cB \pare{\cH}$ s.t. it transforms $\ket{\Gamma}$ into $\ket{\psi}$. Its explicit action would be written as 

\ben
\hspace{-2cm} 
\pare{\id_A \otimes V_{\ket{\psi}}} \ket{\Gamma}=&\sum_{i,j=0}^{d-1}  \alpha_{ij} \pare{\id_A \otimes \op{j}{i} } \pare{ \sum_{k=0}^{d-1} \ket{k} \otimes \ket{k} }
   = \sum_{i,j,k=0}^{d-1} \alpha_{ij} \pare{\ket{k} \otimes \ket{j}} \bracket{i}{k} \nonumber\\
   &= \sum_{i,j,k=0}^{d-q}  \alpha_{ij} \pare{\ket{k} \otimes \ket{j}} \delta_{i,k} 
= \sum_{i,j=0}^{d-1} \alpha_{ij} \ket{i}\otimes \ket{j} = \ket{\psi}.   
\label{eq:opsi}
\een
At this point, we have related the vectors in the extended space $\cH_A \otimes \cH$ to operators acting on $\cH$. This can only be done because the vector $\ket{\Gamma}$ is maximally entangled. We go now back to our extended map $\cV_2$. Its action on $\proj{\Gamma}$ is given by Eq. (\ref{eq:map2}) and as it is a positive map it can be expanded as 

\be
\cV_2\pare{\proj{\Gamma}} =\sum_{l=0}^{d^2-1}\proj{v_l}.
\label{eq:exp}
\ee
with $\ket{v_l}\in\cH_A\otimes \cH$. The vectors $\ket{v_l}$ can be related to operators in $\cH$ as in Eq. (\ref{eq:opsi}).  

\be
\ket{v_l}=\pare{\id_A\otimes\ V_l}\ket{\Gamma}.
\ee
Based on this result we can calculate the product of an arbitrary vector $\ket{i}_A\in\cH_A$ with $\ket{v_l}$.

\be
\bra{i}_A \ket{v_l}=\bra{i}_A \pare{\id_A \otimes V_l} \ket{\Gamma}=V_l \sum_{k=0}^{d-1} \bracket{i}{k}_A \otimes\ket{k}.
\ee
This is the last ingredient we need for the proof. 

We come back to the original question, we want to characterise the map $\cV$. We do so by applying it to an arbitrary  basis element $\op{i}{j}$ of $\cB\pare{\cH}$.


\ben
\hspace{-2.cm} 
\cV\pare{\op{i}{j}} = \pare{ \bra{i}_A \otimes \id_A} \cV_2\pare{\proj{\Gamma}} \pare{\ket{j}_A\otimes\id_A}
= \pare{\bra{i}_A \otimes \id_A} \cor{ \sum_{l=0}^{d^2-1} \proj{v_l} }  \pare{ \ket{j}_A\otimes\id_A } \nonumber\\
= \sum_{l=0}^{d^2-1} \cor{\pare{\bra{i}_A \otimes \id_A}\ket{v_l}} \cor{\bra{v_l}\pare{\ket{j}_A \otimes \id_A}} 
= \sum_{l=0}^{d^2-1} V_l\op{i}{j} V_l.  
\een
As $\op{i}{j}$ is an arbitrary element of a basis any operator can be expanded in this basis. Therefore, it is straightforward to prove that 

\be
\cV \rho=\sum_l^{d^2-l} V^{\dagger}_l\rho V^{\phantom{\dagger}}_l.
\nonumber
\ee

\vspace{0.25 cm}
\noindent
\textbf{End of the proof.}
\vspace{0.5cm}

\noindent
Thanks to Choi's Theorem, we know the general form of CP-maps, but there is still an issue to address. As density matrices should have trace one, we need to require  any physical maps to be also trace-preserving. This requirement gives as a new constraint that completely defines all CPT-maps. This requirement comes from the following theorem. 

\vspace{0.5cm}
\begin{theorem}
Choi-Kraus' Theorem. 

\noindent
A linear map $\cV:B(\cH)\to B(\cH)$ is completely positive and trace-preserving iff it can be expressed as 

\be
\cV\rho=\sum_l V_l^\dagger \rho V^{\phantom{\dagger}}_l
\label{eq:choi2}
\ee
with $V_l\in B(\cH)$ fulfilling  

\be
\sum_l V^{\phantom{\dagger}}_l V_l^\dagger=\id_{\cH}.
\label{eq:krauss}
\ee

\end{theorem}
\vspace{0.5cm}

\noindent
\textbf{Proof.}

\noindent
We have already proved that this is a completely positive map, we only need to prove that it is also trace-preserving and that all trace preserving-maps fulfil Eq. (\ref{eq:krauss}). The `if'  proof is quite simple by applying the cyclic permutations  and linearity properties of the trace operator. 

\be
\Tr\cor{\cV\rho}=\Tr \cor{ \sum_{l=1}^{d^2-1} V^{\phantom{\dagger}}_l\rho V_l^\dagger } = \Tr \cor{ \pare{\sum_{l=1}^{d^2-1} V_l^\dagger V^{\phantom{\dagger}}_l  }\rho } =\Tr\cor{\rho}.
\ee 
We have to prove also that any map in the form (\ref{eq:choi2}) is trace-preserving only if the operators $V_l$ fulfil (\ref{eq:krauss}). We start by stating that if the map is trace-preserving by applying it to an any arbitrary element of a basis of $\cB\pare{\cH}$ we should obtain

\be
\Tr\cor{\cV \pare{\op{i}{j}}}=\Tr \cor{\op{i}{j}}=\delta_{i,j}.
\ee
As the map has a form given by (\ref{eq:choi2}) we can calculate this same trace in an alternative way. 

\ben
\Tr\cor{\cV \pare{\op{i}{j}}} &=& \Tr \cor{ \sum_{l=1}^{d^2-1} V^{\phantom{\dagger}}_l \op{i}{j} V_l^\dagger }  
= \Tr \cor{ \sum_{l=1}^{d^2-1} V_l^\dagger V^{\phantom{\dagger}}_l \op{i}{j}  }  \nonumber \\
&= &\sum_{k}  \bra{k}  \pare{ \sum_{l=1}^{d^2-1} V_l^\dagger V^{\phantom{\dagger}}_l \op{i}{j}  } \ket{k} 
=  \bra{j}   \pare{\sum_{l=1}^{d^2-1} V_l^\dagger V^{\phantom{\dagger}}_l }\ket{i},   
\een  
where $\left\{ \ket{k} \right\}$ is an arbitrary basis of $\cH$. 
As both equalities should be right we obtain 

\be
\bra{j}   \pare{\sum_{l=1}^{d^2-1} V^{\phantom{\dagger}}_l V^\dagger_l }\ket{i} = \delta_{i,j},
\ee
and therefore, the condition (\ref{eq:krauss}) should be fulfilled.

 \vspace{0.25 cm}
\noindent
\textbf{End of the proof.}
\vspace{0.5cm}

\noindent
Operators $V_i$ of a map fulfilling condition (\ref{eq:krauss}) are called {\it Krauss operators}. Because of that, sometimes CPT-maps are also called {\it Krauss maps}, especially when they are presented as a collection of Krauss operators. Both concepts are ubiquitous in quantum information science. Krauss operators can also be time-dependent as long as they fulfil relation (\ref{eq:krauss}) for all times. 

At this point, we already know the form of CPT-maps, but we do not have a master equation, that is a continuous set of differential equations. This means that we know how to perform an arbitrary operation in a system, but we do not have an equation to describe its time evolution. To do so, we need to find a time-independent generator $\cL$ such that 

\be
\frac{d}{dt} \rho\pare{t}= \cL\rho(t),
\label{eq:difeq}
\ee
and therefore our CPT-map could be expressed as $\cV(t)=e^{\cL t}$. The following calculation is about founding the explicit expression of $\cL$. We start by choosing an orthonormal basis of the bounded space of operators $\cB(\cH)$, $\key{F_i}_{i=1}^{d^2}$. To be orthonormal it should satisfy the following condition

\be
\bbracket{F_i}{F_j}\equiv \Tr\cor{F_i^\dagger F_j}=\delta_{i,j}.
\label{eq:orthonormal}
\ee
Without any loss of generality, we select one of the elements of the basis to be proportional to the identity, $F_{d^2}=\frac{1}{\sqrt{d}} \id_{\cH}$. It is trivial to prove that the norm of this element is one, and it is easy to see from Eq. (\ref{eq:orthonormal}) that all the other elements of the basis should have trace zero. 

\be
\Tr\cor{F_i}=0 \qquad \forall i=1,\dots,d^2-1.
\ee
The closure relation of this basis is $\id_{\cB(\cH)}=\sum_i \bproj{F_i}{F_i}$. Therefore, the Krauss operators can be expanded in this basis by using the Fock-Liouville notation

\be
V_l(t)= \sum_{i=1}^{d^2} \bbracket{F_i}{V_l(t)} \kket{F_i}.
\label{eq:kraussexpansion}
\ee
As the map $\cV(t)$ is in the form (\ref{eq:choimap}) we can apply (\ref{eq:kraussexpansion}) to obtain\footnote{ For simplicity, in this discussion we omit the explicit time-dependency of the density matrix.}. 

\be
\hspace{-2cm}
\cV(t) \rho = \sum_l\cor{\sum_{i=1}^{d^2} \bbracket{F_i}{V_l(t)} F_i \;\rho \sum_{j=1}^{d^2} F_j^\dagger \bbracket{V_l(t)}{F_j}} 
=  \sum_{i,j=1}^{d^2} c_{i,j}(t) F_i^{\phantom{\dagger}}\rho F_j^{\dagger},
\ee
where we have absorved the sumation over the Krauss operators in the terms $c_{i,j}(t)= \sum_l \bbracket{F_i}{V_l} \bbracket{V_l}{F_j}$. We go back now to the original problem by applying this expansion into the time-derivative of Eq. (\ref{eq:difeq})

\ben
\frac{d \rho }{dt} &=&\lim_{\Delta t\to0} \frac{1}{\Delta t} \pare{\cV(\Delta t)\rho-\rho} 
= \lim_{\Delta t\to 0} \pare{\sum_{i,j=1}^{d^2} c_{i,j}(\Delta t) F_i^{\phantom{\dagger}} \rho F_j^\dagger - \rho} \nonumber\\
&=& \lim_{\Delta t\to 0} \left( \sum_{i,j=0}^{d^2-1}  c_{i,j}(\Delta t)  F_i^{\phantom{\dagger}}\rho F_j^\dagger + \sum_{i=1}^{d^2-1} c_{i,d^2} F_i^{\phantom{\dagger}}\rho F_{d^2}^\dagger  \right. \nonumber\\
&&\left. +  \sum_{j=1}^{d^2-1} c_{d^2,j} (\Delta t) F_{d^2}^{\phantom{\dagger}} \rho F_j^\dagger + c_{d^2,d^2}(\Delta t)  F_{d^2}^{\phantom{\dagger}} \rho F_{d^2}^\dagger - \rho \right),
\een
where we have separated the summations to take into account that $F_{d^2}=\frac{1}{\sqrt{d}}\id_{\cH}$.  By using this property this equation simplifies to

 \ben
\frac{d \rho}{dt} &=&\lim_{\Delta t\to0} \frac{1}{\Delta t} \left(  \sum_{i,j=1}^{d^2-1} c_{i,j}(\Delta t) F_i^{\phantom{\dagger}} \rho F_j^\dagger + \frac{1}{\sqrt{d}}  \sum_{i=1}^{d^2-1} 
 c_{i,d^2}(\Delta t) F_i^{\phantom{\dagger}}\rho \right. \nonumber\\
 &+&\left.  \frac{1}{\sqrt{d}} \sum_{j=1}^{d^2-1} c_{d^2,j} (\Delta t) \rho F_j^\dagger + \frac{1}{d} c_{d^2,d^2}(\Delta t) \rho-\rho \right).
 \label{eq:derivative2}
\een
The next step is to eliminate the explicit dependence with time. To do so, we define  new constants to absorb all the time intervals. 

\ben
g_{i,j}&\equiv& \lim_{\Delta t\to 0} \frac{c_{i,j} (\Delta t) }{\Delta t} \qquad (i,j<d^2), \nonumber\\
g_{i,d^2}&\equiv& \lim_{\Delta t \to 0} \frac{c_{i,d^2} (\Delta t) }{\Delta t} \qquad (i<d^2), \nonumber\\
g_{d^2,j}&\equiv& \lim_{\Delta t \to 0} \frac{c_{d^2,j} (\Delta t) }{\Delta t} \qquad (j<d^2), \\
g_{d^2,d^2}&\equiv& \lim_{\Delta t\to 0} \frac{c_{d^2,d^2}(\Delta t)-d}{\Delta t}.  \nonumber
\een 
Introducing these coefficients in Eq (\ref{eq:derivative2}) we obtain an equation with no explicit dependence in time. 

\ben
\frac{d \rho}{dt} = \sum_{i,j=1}^{d^2-1} g_{i,j} F_i \rho F_j^\dagger + \frac{1}{\sqrt{d}} \sum_{i=1}^{d^2-1} g_{i,d^2} F_i \rho + \frac{1}{\sqrt{d}} \sum_{j=1}^{d^2-1} g_{d^2,j}  \rho F_j^\dagger + \frac{g_{d^2,d^2}}{d} \rho.\nonumber \\ 
\label{eq:derivative3}
\een
As we are already summing up over all the Krauss operators it is useful to define a new operator

\be
F\equiv \frac{1}{\sqrt{d}} \sum_{i=1}^{d^2-1} g_{i,d^2} F_i.
\ee
Applying it to Eq. (\ref{eq:derivative2}).

\be
\frac{d \rho}{dt} = \sum_{i,j=1}^{d^2-1} g_{i,j} F_i \rho F_j^\dagger + F \rho + \rho F^\dagger + \frac{g_{d^2,d^2}}{d} \rho.
\label{eq:derivative4}
\ee
At this point, we want to separate the dynamics of the density matrix into a Hermitian (equivalent to von Neunmann equation) and an incoherent part. We split the operator $F$ in two to obtain a Hermitian and anti-Hermitian part.

\be
F=\frac{F+F^\dagger}{2} + i\frac{F-F^\dagger}{2i} \equiv G-iH,
\ee
where we have used the notation $H$ for the Hermitian part for obvious reasons. If we take this definition to Eq. (\ref{eq:derivative4}) we obtain

\be
\frac{d \rho}{dt} = g_{i,j} F_i \rho F_j^\dagger + \key{G,\rho} - i \cor{H,\rho} + \frac{g_{d^2,d^2}}{d} \rho.
\ee
We define now the last operator for this proof,  $G_2\equiv G+\frac{g_{d^2,d^2}}{2d}$, and the expression of the  time derivative leads to 

\be
\frac{d \rho}{dt} = \sum_{i,j=1}^{d^2-1} g_{i,j} F_i \rho F_j^\dagger + \key{G_2,\rho} -i\cor{H,\rho}. 
\ee
Until now we have imposed the complete positivity of the map, as we have required it to be written in terms of Krauss maps, but we have not used the trace-preserving property. We impose now this property, and by using the cyclic property of the trace, we obtain a new condition

\be
\Tr\cor{\frac{d \rho}{dt}}=\Tr\cor{ \sum_{i,j=1}^{d^2-1} F_j^\dagger F_i \rho + 2 G_2 \rho }=0.
\ee
Therefore, $G_2$ should fulfil 

\be
G_2=\frac{1}{2} \sum_{i,j=1}^{d^2-1} g_{i,j} F_j^\dagger F_i\rho.
\ee
By applying this condition, we arrive at the Lindblad master equation  

\be
 \frac{d \rho}{dt}= -i\cor{H,\rho} + \sum_{i,j=1}^{d^2-1} g_{i,j} \pare{ F_i^{\phantom{\dagger}}\rho F_j^\dagger - \frac{1}{2} \key{F_j^\dagger F_i^{\phantom{\dagger}},\rho }}.
\ee
Finally, by definition the coefficients $g_{i,j}$ can be arranged to form a Hermitian, and therefore diagonalisable, matrix. By diagonalising it, we obtain the diagonal form of the Lindblad master equation. 

\be
\frac{d}{dt} \rho= -i \cor{H,\rho} + \sum_{k} \Gamma_k \pare{ L_k^{\phantom{\dagger}} \rho L_k^\dagger - \frac{1}{2} \key{L_k^{\phantom{\dagger}} L_k^\dagger,\rho}} \equiv \cL \rho.
\label{eq:lindblad}
\ee

\subsection{Properties of the Lindblad Master Equation}


Some interesting properties of the Lindblad equation are: 

\begin{itemize}

\item Under a Lindblad dynamics, if all the jump operators are Hermitian, the purity of a system fulfils  $\frac{d}{dt}\pare{ \Tr \cor{\rho^2 }} \le 0$.  The proof is given in \ref{sec:purity}.

\item The Lindblad Master Equation is invariant under unitary transformations of the jump operators

\be
\sqrt{\Gamma_i} L_i\to \sqrt{\Gamma'_i} L_i'= \sum_j  v_{ij} \sqrt{\Gamma_j} L_j,
\ee
with $v$ representing a unitary matrix. It is also invariant under inhomogeneous transformations in the form 

\ben
L_i &\to& L'_i= L_i + a_i \nonumber\\
H&\to& H'=H+\frac{1}{2i} \sum_j \Gamma_j \pare{a_j^* A_j - a_j A_j^\dagger }+ b,
\een
where $a_i \in \mathbb{C}$ and $b \in \mathbb{R}$. The proof of this can be found in Ref. \cite{breuer_02} (Section 3). 

\item Thanks to the previous properties it is possible to find traceless jump operators without loss of generality.

\end{itemize}

\newpage

\vspace{0.5cm}
\bc
\framebox[15.5cm][l]{
\begin{minipage}[l]{15cm}
\vspace{0.25cm} 
{\bf Box 6. A master equation for a two-level system with decay.}
\vspace{0.25cm} 

Continuing our example of a two-level atom, we can make it more realistic by including the possibility of atom decay by the emission of a photon. This emission happens due to the interaction of the atom with the surrounding vacuum state\footnote{This is why atoms decay.}. The complete quantum system would be in this case the `atom+vacuum' system and its time evolution should be given by the von Neumann equation (\ref{eq:vne}), where $H$ represents the total `atom+vacuum' Hamiltonian. This system belongs to an infinite-dimension Hilbert space, as the radiation field has infinite modes. If we are interested only in the time dependence state of the atom, we can derive a Markovian master equation for the reduced density matrix of the atom (see for instance Refs. \cite{breuer_02,gardiner_00}).  The master equation we will study is 

\ben
\frac{d}{dt}\rho(t) = -i\cor{H,\rho} + \Gamma  \left( \sigma^- \rho \sigma^+ -\frac{1}{2} \left\{\sigma^+ \sigma^-,\rho \right\}  \right),
\label{eq:lindtotal}
\een
where $\Gamma$ is the coupling between the atom and the vacuum.

In the Fock-Liouvillian space  (following the same ordering as in Eq. (\ref{eq:denmat})) the Liouvillian corresponding to evolution (\ref{eq:lindtotal}) is

\be
\cL= 
\left(
\begin{array}{cccc}
0 & i \Omega  & -i\Omega & \Gamma  \\
i\Omega &  -i E - \frac{\Gamma}{2}  & 0 &  -i\Omega\\
-i\Omega & 0  & -i E  -\frac{\Gamma}{2} & i\Omega \\
0 &  -i\Omega  &  i\Omega & -\Gamma  \\
\end{array}
\right).
\label{eq:liou2}
\ee

Expressing explicitly the set of differential equations we obtain

\ben
\dot{\rho}_{00} &=  & i \Omega \rho_{01} -i\Omega  \rho_{10} + \Gamma  \rho_{11} \nonumber \\
\dot{\rho}_{01} &=& i\Omega  \rho_{00}  - \pare{ iE - \frac{\Gamma}{2} }  \rho_{01}    -i\Omega  \rho_{11} \nonumber\\
\dot{\rho}_{10} &=& -i\Omega \rho_{00}  \pare{ -i E - \frac{\Gamma}{2}} \rho_{10} + i\Omega \rho_{11} \\
\dot{\rho}_{10} &=&  -i\Omega   \rho_{01} +  i\Omega  \rho_{10} -\Gamma   \rho_{11} \nonumber
\een

\vspace{0.25cm}

\label{minipage5}
\end{minipage}
}
\ec
\vspace{0.5cm}

\newpage
\section{Resolution of the Lindblad Master Equation}
\label{sec:resolution}

\subsection{Integration}

To calculate the time evolution of a system determined by a Master Equation in the form (\ref{eq:lindtotal}) we need to solve a set of equations with as many equations as the dimension of the density matrix. In our example, this means to solve a 4 variable set of equations, but the dimension of the problem increases exponentially with the system size. Because of this, for bigger systems techniques for dimension reduction are required. 

To solve systems of partial differential equations there are several canonical algorithms. This can be done analytically only for a few simple systems and by using sophisticated techniques as damping bases \cite{briegel:pra93}. In most cases, we have to rely on numerical approximated methods. One of the most popular approaches is the $4^{th}$-order Runge-Kutta algorithm (see, for instance, \cite{numericalrecipes} for an explanation of the algorithm). By integrating the equations of motion, we can calculate the density matrix at any time $t$. 

The steady-state of a system can be obtained by evolving it for a long time $\pare{t \rightarrow \infty}$. Unfortunately, this method presents two difficulties. First, if the dimension of the system is big, the computing time would be huge. This means that for systems beyond a few qubits, it will  take too long to reach the steady-state. Even worse is the problem of stability of the algorithms for integrating differential equations. Due to small errors in the calculation of derivatives by the use of finite differences, the trace of the density matrix may not be constantly equal to one. This error accumulates during the propagation of the state, giving non-physical results after a finite time. One solution to this problem is the use of algorithms specifically designed to preserve the trace, as Crank-Nicholson algorithm \cite{goldberg:ajp67}. The problem with this kind of algorithms is that they consume more computational power than Runge-Kutta, and therefore they are not useful to calculate the long-time behaviour of big systems. An analysis of different methods and their advantages and disadvantages can be found at Ref. \cite{riesch:jcp19}.


\vspace{.35cm}

\bc
\framebox[15.5cm][l]{
\begin{minipage}[l]{15cm}
\vspace{0.25cm} 
{\bf Box 7. Time dependency of the two-level system with decay.}

\vspace{0.25cm} 

In this box we show some results of solving Eq (\ref{eq:lindtotal}) and calculating the density matrix as a function of time. A Mathematica notebook solving this problem can be found at \cite{notebook}. To illustrate the time behaviour of this system, we calculate the evolution for different state parameters. In all cases, we start with an initial state that represents the state being excited $\rho_{11}=1$, with no coherence between different states, meaning $\rho_{01}=\rho_{10}=0$.  If the decay parameter $\Gamma$ is equal to zero, the problem reduces to solve von Neumann equation, and the result is displayed in Figure \ref{figure1}. The other extreme case would be a system with no coherent dynamics ($\Omega=0$) but with decay. In this case, we observe an exponential decay of the population of the excited state. Finally, we can calculate the dynamics of a system with both coherent driving and decay. In this case, both behaviours coexist, and there are oscillations and decay.

\bc
\includegraphics[scale=.5]{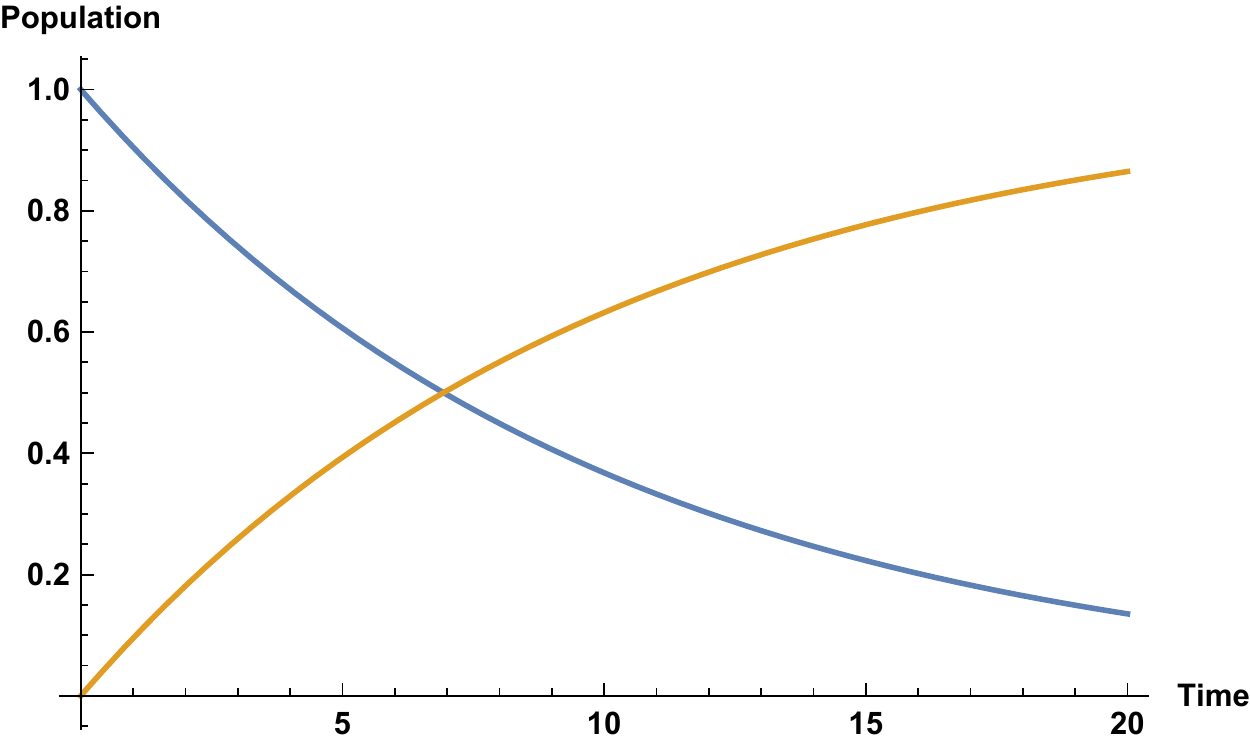}
\includegraphics[scale=.5]{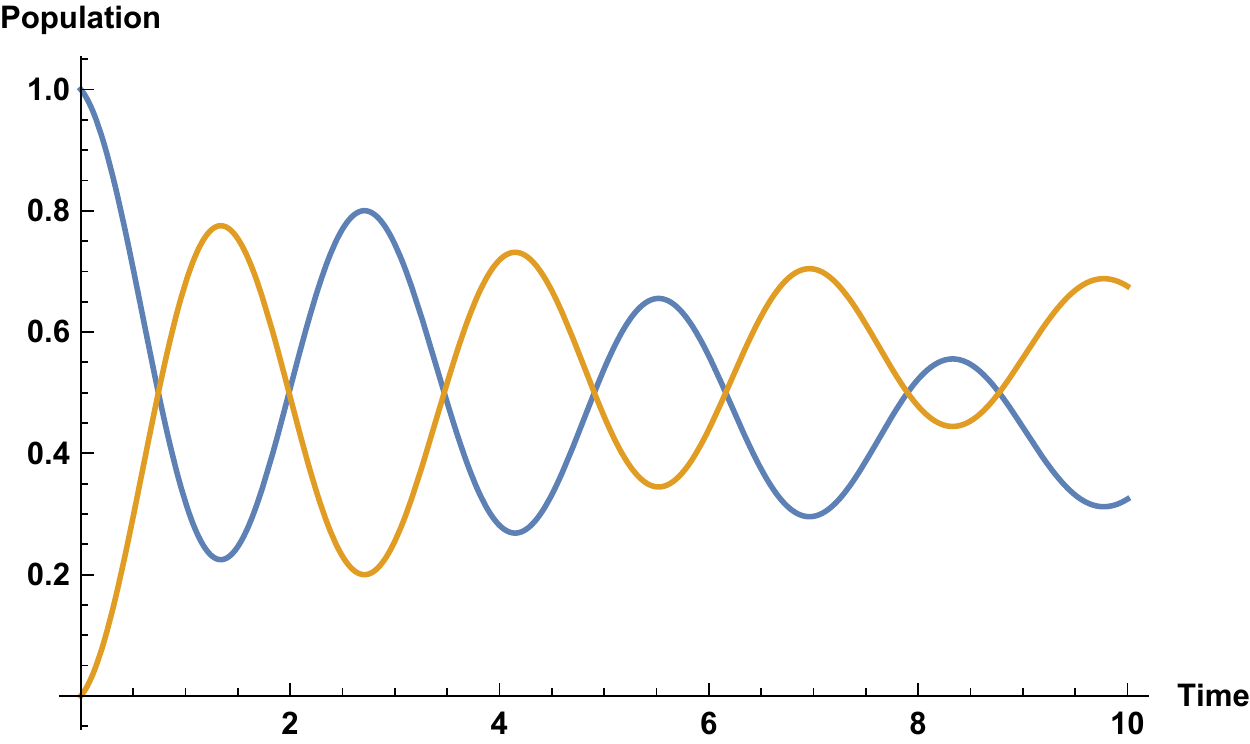}
\captionof{figure}{Left: Population dynamics under a pure incoherent dynamics ($\Gamma=0.1,\; n=1,\; \Omega=0,\; E=1$). Right: Population dynamics under both coherent and incoherent dynamics ($\Gamma=0.1,\; n=1,\; \Omega=1,\; E=1)$.  In both the blue lines represent $\rho_{11}$ and the orange one $\rho_{00}$.}
\ec

\vspace{0.1cm}
\end{minipage}
\label{minipage6}
}
\ec

\newpage

\subsection{Diagonalisation}

As we have discussed before, in the Fock-Liouville space the Liouvillian corresponds to a complex matrix (in general complex, non-hermitian, and non-symmetric). By diagonalising it we can calculate both the time-dependent and the steady-state of the density matrices. For most purposes, in the short time regime integrating the differential equations may be more efficient than diagonalising. This is due to the high dimensionality of the Liouvillian that makes the diagonalisation process very costly in computing power. On the other hand, in order to calculate the steady-state, the  diagonalisation is the most used method due to the problems of integrating the equation of motions discussed in the previous section. 

Let see first how we use diagonalisation to calculate the time evolution of a system. As the Liouvillian matrix is non-Hermitian, we cannot apply the spectral theorem to it, and  it may have different left and right eigenvectors. For a specific eigenvalue $\Lambda_i$ we can obtain the eigenvectors $\kket{\Lambda_i^R}$ and $\kket{\Lambda_i^L}$ s. t. 

\ben
\hspace{2cm}
\tilde{\cL} \; \kket{\Lambda_i^R} = \Lambda_i \kket{\Lambda_i^R} \nonumber\\
\hspace{2cm}
\bbra{\Lambda_i^L}\; \tilde{\cL} =  \Lambda_i \bbra{\Lambda_i^L} 
\een
An arbitrary system can be expanded in the eigenbasis of $\tilde{\cL}$ as \cite{thingna:sr16,gardiner_00} 

\be
\kket{\rho(0)}= \sum_i \kket{\Lambda_i^R} \bbracket{\Lambda_i^L}{\rho(0)}.
\ee
Therefore, the state of the system at a time $t$ can be calculated in the form 

\be
\kket{\rho(t)}= \sum_i  e^{\Lambda_i t} \kket{\Lambda_i^R}    \bbracket{\Lambda_i^L}{\rho(0)}.
\ee
Note that in this case to calculate the state a time $t$ we do not need to integrate into the interval $\cor{0,t}$, as we have to do if we use a numerical solution of the differential set of equations. This is an advantage when we want to calculate long-time behaviour. Furthermore, to calculate the steady-state of a system, we can look to the eigenvector that has zero eigenvalue, as this is the only one that survives when $t\to\infty$. 

For any finite system, Evans' Theorem ensures the existence of at least one zero eigenvalue of the Liouvillian matrix \cite{evans:cmp77,evans:jfa79}. The eigenvector corresponding to this zero eigenvalue would be the steady-state of the system. In exceptional cases, a Liouvillian can present more than one zero eigenvalues due to the presence of symmetry in the system \cite{buca:njp12,manzano:prb14,manzano:av18}. This is a non-generic case, and for most purposes, we can assume the existence of a unique fixed point in the dynamics of the system. Therefore, diagonalising can be used to calculate the steady-state without calculating the full evolution of the system. This can be done even analytically for small systems, and when numerical approaches are required this technique gives better precision than integrating the equations of motion. The spectrum of Liouvillian superoperators has been analysed in several recent papers \cite{albert:pra14,thingna:sr16}. 

\newpage
\vspace{0.5cm}
\framebox[15.5cm][l]{
\begin{minipage}[l]{15cm}
\vspace{0.25cm} 

{\bf Box 8. Spectrum-analysis of the Liouvillian for the two-level system with decay.}

Here we diagonalise  (\ref{eq:liou2}) and obtain its steady state. A Mathematica notebook solving this problem can be downloaded from \cite{notebook}. This specific case is straightforward to diagonalize as the dimension of the system is very low. We obtain $4$ different eigenvalues, two of them are real while the other two form a conjugated pair.  Figure \ref{fig:spectrum} sisplays the spectrum of the superoperator $\cL$ given in (\ref{eq:liou2}).

\bc
\includegraphics[scale=.9]{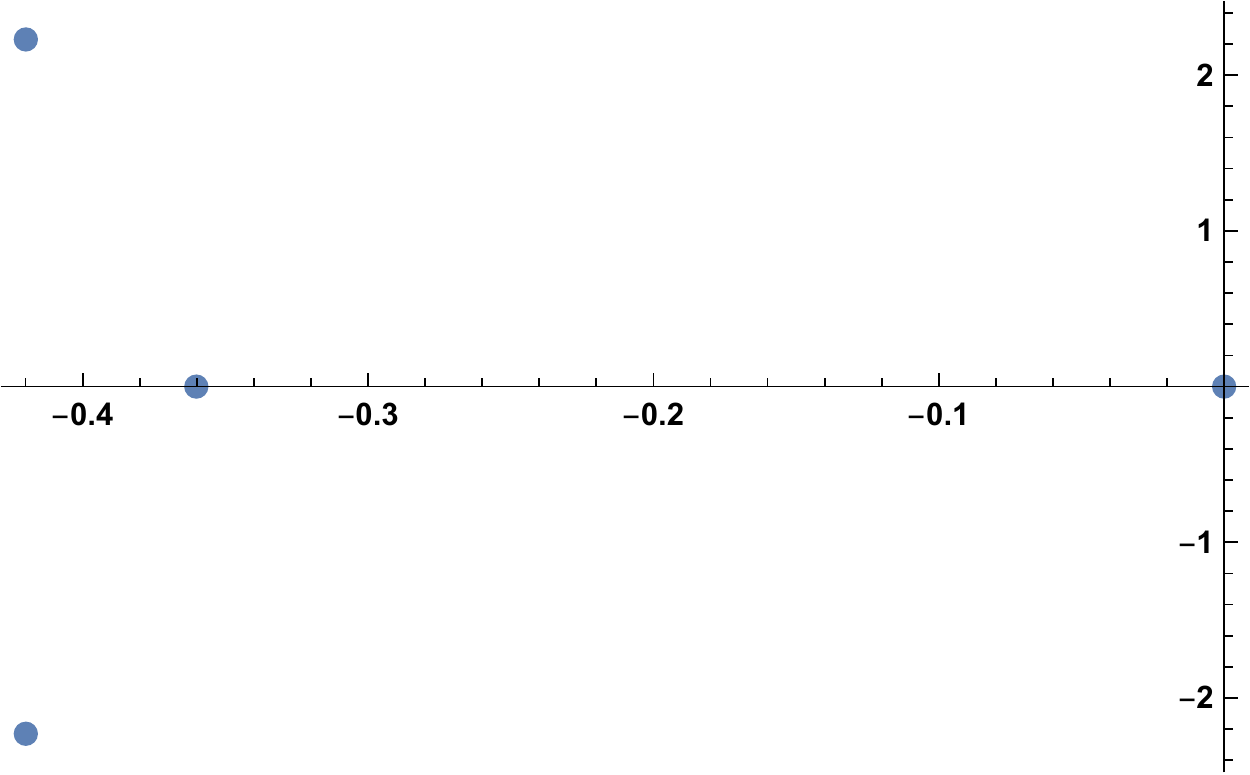}
\captionof{figure}{Spectrum of the Liouvillian matrix given by (\ref{eq:liou2}) for the general case of both coherent and incoherent dynamics ($\Gamma=0.2,\; n=1,\; \Omega=0,\; E=1$).}
\label{fig:spectrum}
\ec

As there only one zero eigenvalue we can conclude that there is only one steady-state, and any initial density matrix will evolve to it after an infinite-time evolution. By selecting the right eigenvector corresponding to the zero-eigenvalue and normalizing it we obtain the density matrix. This can be done even analytically. The result is the matrix:

\be
\rho_{SS}=
\left(
\begin{array}{cc}
 \frac{(1+n) \left(4\, E^2+(\Gamma +2 n\, \Gamma )^2\right)+4 (1+2 n) \Omega ^2}{(1+2 n) \left(4 \,E^2+(\Gamma +2 n \,\Gamma )^2+8 \Omega^2\right)} & \frac{2 (-2 \,E-i (\Gamma +2 n \Gamma )) \Omega }{(1+2 n) \left(4 \,E^2+(\Gamma +2 n \,\Gamma )^2+8 \,\Omega ^2\right)} \\
 \frac{2 (-2 \,E+i (\Gamma +2 n\, \Gamma )) \Omega }{(1+2 n) \left(4\, E^2+(\Gamma +2 n\, \Gamma )^2+8 \Omega ^2\right)} & \frac{n \left(4E^2+(\Gamma +2 n \Gamma )^2\right)+4 (1+2 n) \Omega ^2}{(1+2 n) \left(4\, E^2+(\Gamma +2 n \Gamma )^2+8\, \Omega ^2\right)} \\
\end{array}
\right)
\ee

\vspace{0.25cm}
\vspace{0.25cm} 

\end{minipage}
}

\section{Acknowledgements}
 The author wants to acknowledge the Spanish Ministry and the Agencia Espa{\~n}ola de Investigaci{\'o}n (AEI) for financial support under grant
FIS2017-84256-P (FEDER funds).



\appendix
\setcounter{section}{0}

\newpage
\section{ Proof of  $\frac{d}{dt} \Tr \cor{\rho^2} \le 0$} 
\label{sec:purity}

In this appendix we proof that under the Lindblad dynamics given by Eq. (\ref{eq:lindblad}) the purity of a density matrix fulfils that $\frac{d}{dt} \Tr \cor{\rho^2} \le 0$ if all the jump operators of the Lindblad dynamics are Hermitian.

We start just by interchanging the trace and the derivative. As the trace is a linear operation it commutes with the derivation, and we have

\be
\frac{d}{dt}\pare{ \Tr\cor{\rho^2}} = \Tr\cor{ \frac{ d \rho^2}{dt}} = \Tr\cor{ 2 \rho \dot{\rho}  }, 
\label{eq:lpurity1}
\ee
where we have used the cyclic property of the trace operator\footnote{This property is used along all the demonstration without explicitly mentioning it.}. By inserting the Lindblad Eq. (\ref{eq:lindblad}) into the r.h.s of (\ref{eq:lpurity1}) we obtain 

\ben
\frac{d}{dt}\pare{ \Tr\cor{\rho^2}} &=& - \frac{i}{\hbar} \Tr \cor{\pare{ 2\rho \pare{H\rho -\rho H}} }  \nonumber \\
&+&  2 \sum_k \Gamma_k \Tr \cor{ \rho \, L_k^{\phantom{\dagger}} \, \rho \, L_k^\dagger } -  2\sum_k \Gamma_k \Tr \cor{ \rho^2 L_k^\dagger L_k^{\phantom{\dagger}}  }.
\een
The first term is zero. Therefore, the inequality we want to prove  becomes equivalent to 

\be
\sum_k \Gamma_k \Tr \cor{ \rho\, L_k^{\phantom{\dagger}}\, \rho\, L_k^\dagger } \le \sum_k \Gamma_k \Tr\cor{ \rho^2  L_k^\dagger L_k^{\phantom{\dagger}}}
\label{eq:lpurity2}
\ee
As the density matrix is Hermitian we can diagonalize it to obtain its eigenvalues ($\Lambda_i \in \mathbb{R}$) and its corresponding eigenvectors ($\ket{\Lambda_i}$). The density matrix is diagonal in its own eigenbasis and can be expressed as\footnote{This eigenbasis changes with time, of course, but the proof is valid as the inequality should be fulfilled at any time.}

\be
\rho\to \tilde{\rho}=\sum_i \Lambda_i \proj{\Lambda_i},
\ee
where we assume an ordering of the eigenvalues in the form $\Lambda_0\ge \Lambda_1 \ge \cdots \ge \Lambda_d$.

We rename the jump operators in this basis as $\tilde{L}_i$ a. Expanding each term of the inequality (\ref{eq:lpurity2}) in this basis we obtain 

\ben
\sum_k \Gamma_k  \Tr \cor{ \rho\, L_k\, \rho\, L_k^\dagger } = \sum_k \Gamma_k \Tr \cor{ \pare{ \sum_i \Lambda_i \proj{\Lambda_i} } \tilde{L}_k  \pare{\sum_j  \Lambda_j \proj{\Lambda_j} } \tilde{L}_k } \nonumber\\
=\sum_k \Gamma_k  \sum_{i,j} \Lambda_i \Lambda_j \Tr  \cor{ \tilde{L}_k^\dagger \proj{\Lambda_i} \tilde{L}_k \proj{\Lambda_j}}  
=\sum_k  \Gamma_k  \sum_{i,j} \Lambda_i \Lambda_j \Tr  \cor{ \left| \bra{\Lambda_i} \tilde{L}_k \ket{\Lambda_j}  \right|^2 }\nonumber \\
=  \sum_k \Gamma_k \sum_{i,j} \Lambda_i \Lambda_j x_{ij}^{(k)}, 
\een  
where we have introduced the oefficients $x_{ij}^{(k)} \equiv  \left| \bra{\Lambda_i} \tilde{L}_k \ket{\Lambda_j}  \right|^2 $. As the operators $L_k$ are Hermitian these coefficients fulfil  $x_{ij}^{(k)}=x_{ji}^{(k)}$

The second term is expanded as 

\ben
\sum_k  \Gamma_k \Tr \cor{ \rho^2 L_k^\dagger L_k  } = \sum_k  \Gamma_k \Tr \cor{ \pare{\sum_i \Lambda_i \proj{\Lambda_i} } \pare{\sum_j \Lambda_j \proj{\Lambda_j}} \tilde{L}^\dagger_k \tilde{L}_k}  \nonumber\\
= \sum_k  \Gamma_k \sum_{ij} \Lambda_i \Lambda_j  \Tr \cor{   \tilde{L}_k \op{\Lambda_i}{\Lambda_j} \tilde{L}_k^\dagger\bracket{\Lambda_i}{\Lambda_j}  }  
= \sum_k  \Gamma_k \sum_i  \Lambda_i^2 \Tr \cor{ \tilde{L}_k \proj{\Lambda_i} \tilde{L}_k^\dagger\; } \nonumber\\
= \sum_k  \Gamma_k \sum_i \Lambda_i^2 \Tr \cor{ \tilde{L}_k \proj{\Lambda_i} \tilde{L}_k^\dagger  \pare{\sum_j \proj{\Lambda_j}} } \nonumber\\
= \sum_k  \Gamma_k \sum_{ij} \Lambda_i^2 \Tr \cor{ \bra{\Lambda_j}\tilde{L}_k \ket{\Lambda_i}  + \bra{\Lambda_i}\tilde{L}_k \ket{\Lambda_j}  }  
= \sum_k  \Gamma_k \sum_{ij} \Lambda_i^2 x_{ij}, 
\een
where we have used the closure relation in the density matrix eigenbasis,  $\id_{\cH}=\sum_j \proj{\Lambda_j}$. The inequality can be written now as 

\be
\sum_k  \Gamma_k \sum_{ij} \Lambda_i \Lambda_j x_{ij} \le \sum_k   \Gamma_k \sum_{ij} \Lambda_i^2 x_{ij}.
\ee
As $x_{ij}=x_{ji}$ we can re-order the $ij$ sum in the following way 

\be
\sum_k  \Gamma_k \sum_i \pare{ \sum_{j\le i} 2 \Lambda_i \Lambda_j x_{ij}^{(k)} + \Lambda_i^2 x_{ii}^{(k)} } \le \sum_k  \Gamma_k\sum_i  \pare{ \sum_{j<i} \pare{ \Lambda_i^2 + \Lambda_j^2}  x_{ij}^{(k)} + \Lambda_i^2 x_{ii}^{(k)} }.
\label{eq:lpurity3}
\ee
Therefore, we can reduce the proof of this inequality to the proof of a set of inequalities 

\be
2\Lambda_i\Lambda_j x_{ij}^{(k)} \le \pare{\Lambda_i^2 + \Lambda_j^2} x_{ij}^{(k)} \qquad \forall \pare{k,i,j}.
\label{eq:lpurity4}
\ee
It is obvious that (\ref{eq:lpurity4}) $\Rightarrow$ (\ref{eq:lpurity3}) (but not the other way around). The inequalities (\ref{eq:lpurity4}) are easily proved just by taking into account that  $x_{ij}^{(k)} \ge 0$ and applying the Triangular Inequality.

\end{document}